\documentclass[sigconf]{acmart}

\usepackage{booktabs} % For formal tables
\usepackage[ruled,noend,noline,linesnumbered]{algorithm2e} % For algorithms
\usepackage[utf8]{inputenc}
\usepackage[T1]{fontenc}
\usepackage{graphicx}
\usepackage{color}
\usepackage{todonotes}
\usepackage{balance}
\usepackage{tikzsymbols}

\setcopyright{none}
\renewcommand\footnotetextcopyrightpermission[1]{} % removes footnote with conference information in first column
\begin{document}

% Some global commands
\newcommand{\eat}[1]{}

\definecolor{orange}{RGB}{255,102,0}
\definecolor{blue}{RGB}{0,153,255}

\newcommand{\osk}[1]{\todo[inline,color=orange!40,author=Omar]{ #1}}
\newcommand{\hr}[1]{\todo[inline,color=blue!40,author=Hanna]{ #1}}
\newcommand{\tho}[1]{\todo[inline,color=blue!40,author=Þórhildur]{ #1}}
\newcommand{\bj}[1]{\todo[inline,color=green!40,author=Bj\"{o}rn]{ #1}}
\newcommand{\gtg}[1]{\todo[inline,color=purple!40,author=Gylfi]{ #1}}
\newcommand{\jz}[1]{\todo[inline,color=yellow!40,author=Honza]{ #1}}
\newcommand{\sr}[1]{\todo[inline,color=magenta!40,author=Stevan]{ #1}}
\newcommand{\la}[1]{\todo[inline,color=cyan!40,author=Laurent]{ #1}}
\newcommand{\mw}[1]{\todo[inline,color=red!40,author=Marcel]{ #1}}

\newcommand{\req}[1]{\textbf{R#1}}
\newcommand{\reqsrt}{\req{1}}
\newcommand{\reqpqe}{\req{2}}
\newcommand{\reqkfn}{\req{3}}

% Title portion
\title{Exquisitor: Interactive Learning at Large}
\author{Björn Þór Jónsson}
\affiliation{\institution{IT Univ. of Copenhagen, Denmark}}
\affiliation{\institution{Reykjavik University, Iceland}}
\email{bjorn@itu.dk}
\author{Omar Shahbaz Khan}
\affiliation{\institution{IT University of Copenhagen}
\city{Copenhagen}\country{Denmark}}
\email{omsh@itu.dk}
\author{Hanna Ragnarsdóttir}
\affiliation{\institution{Reykjavik University}
\city{Reykjavik}\country{Iceland}}
\email{hannar15@ru.is}
\author{Þórhildur Þorleiksdóttir}
\affiliation{\institution{Reykjavik University}
\city{Reykjavik}\country{Iceland}}
\email{thorhildurt15@ru.is}
\author{Jan Zahálka}
\affiliation{\institution{bohem.ai}
\city{Prague}\country{Czech Republic}}
\email{jan.zahalka@bohem.ai}
\author{Stevan Rudinac}
\affiliation{\institution{University of Amsterdam}
\city{Amsterdam}\country{Netherlands}}
\email{s.rudinac@uva.nl}
\author{Gylfi Þór Guðmundsson}
\affiliation{\institution{Reykjavik University}
\city{Reykjavik}\country{Iceland}}
\email{gylfig@ru.is}
\author{Laurent Amsaleg}
\affiliation{\institution{CNRS--IRISA}
\city{Rennes}\country{France}}
\email{laurent.amsaleg@irisa.fr}
\author{Marcel Worring}
\affiliation{\institution{University of Amsterdam}
\city{Amsterdam}\country{Netherlands}}
\email{m.worring@uva.nl}

\begin{abstract}
Increasing scale is a dominant trend in today's multimedia collections, which especially impacts interactive applications. To facilitate interactive exploration of large multimedia collections, new approaches are needed that are capable of learning on the fly new analytic categories based on the visual and textual content. To facilitate general use on standard desktops, laptops, and mobile devices, they must furthermore work with limited computing resources. We present Exquisitor, a highly scalable interactive learning approach, capable of intelligent exploration of the large-scale YFCC100M image collection with extremely efficient responses from the interactive classifier. Based on relevance feedback from the user on previously suggested items, Exquisitor uses semantic features, extracted from both visual and text attributes, to suggest relevant media items to the user. Exquisitor builds upon the state of the art in large-scale data representation, compression and indexing, introducing a cluster-based retrieval mechanism that facilitates the efficient suggestions. With Exquisitor, each interaction round over the full YFCC100M collection is completed in less than 0.3 seconds using a single CPU core. That is 4x less time using 16x smaller computational resources than the most efficient state-of-the-art method, with a positive impact on result quality. These results open up many interesting research avenues, both for exploration of industry-scale media collections and for media exploration on mobile devices.
\end{abstract}

% The code below should be generated by the tool at
% http://dl.acm.org/ccs.cfm
% Please copy and paste the code instead of the example below.
%
%\begin{CCSXML}
%<ccs2012>
%<concept>
%<concept_id>10002951.10003317.10003371.10003386</concept_id>
%<concept_desc>Information systems~Multimedia and multimodal retrieval</concept_desc>
%<concept_significance>500</concept_significance>
%</concept>
%<concept>
%<concept_id>10002951.10003227.10003251.10003253</concept_id>
%<concept_desc>Information systems~Multimedia databases</concept_desc>
%<concept_significance>500</concept_significance>
%</concept>
%</ccs2012>
%\end{CCSXML}
%
%\ccsdesc[500]{Information systems~Multimedia and multimodal retrieval}
%\ccsdesc[500]{Information systems~Multimedia databases}


\maketitle

% The default list of authors is too long for headers.
%\renewcommand{\shortauthors}{B. Þ. Jónsson et al.}
%\renewcommand{\shorttitle}{High-Performance Interactive Multimodal Learning}

%%%%% Paper Outline %%%%%%%%%%%%%%%%%%%%%%%%%
% Front + intro: 2 pages
% Incremental multimodal learning: 1 (-) page
% High-Dimensional indexing: 1 1/2 (-) pages
% New method: 3 / 4 page
% Experiments:  2 (+) pages
% Conclusions: 1/4 (+) page
% References: Excluded from page limit
% Total: 7 1/2 (-) pages (which is maximum)
%%%%%%%%%%%%%%%%%%%%%%%%%%%%%%%%%%%%%%%%%%%%%

%\osk{Use the $\backslash{}$osk$\left\{\right\}$ command to make comments}
%\hr{Use the $\backslash{}$hr$\left\{\right\}$ command to make comments}
%\tho{Use the $\backslash{}$tho$\left\{\right\}$ command to make comments}
%\bj{Use the $\backslash{}$bj$\left\{\right\}$ command to make comments}
%\gtg{Use the $\backslash{}$gtg$\left\{\right\}$ command to make comments}
%\jz{Use the $\backslash{}$jz$\left\{\right\}$ command to make comments}
%\sr{Use the $\backslash{}$sr$\left\{\right\}$ command to make comments}
%\la{Use the $\backslash{}$la$\left\{\right\}$ command to make comments}
%\mw{Use the $\backslash{}$mw$\left\{\right\}$ command to make comments}

\section{Introduction}
\label{sec:intro}

%The assumption is that everybody knows what we are talking about if we say multimedia analytics. In fact, we typically use it for multimedia analysis + visual analytics while in this paper it is interactive multimodal learning, so more explanation needed. Getting insight is too vague a term. Would be good to add in the intro terms (and references) considering relevance feedback like methods. In the text we implicitly also say that it would be suited for recommendation, but that is not clear in the beginning. The abstract and intro should clearly set you apart from automatic analysis, what are the applications where fully automatic analysis is not feasible and current methods are not accurate enough.

A dominant trend in multimedia applications for industry and society today is the ever-growing scale of media collections. 
As the general public has been given tools for unprecedented media production, storage and sharing, media generation and consumption have exceeded all expectations. 
Furthermore, upcoming multimedia applications in countless domains, from smart urban spaces and business intelligence to health and wellness, lifelogging, and entertainment, increasingly require joint modelling of visual content and text.
This  vastly increases the number of items that must be dealt with, making scalability an even greater concern \cite{10.1007/978-3-319-73603-7_51}. Underlining the importance of collection scale, the multimedia research community  created the YFCC100M collection, with associated calls for arms to tackle scalability of multimedia applications~\cite{Thomee:2016:YND:2886013.2812802}.

\begin{figure}
\includegraphics[width=\columnwidth]{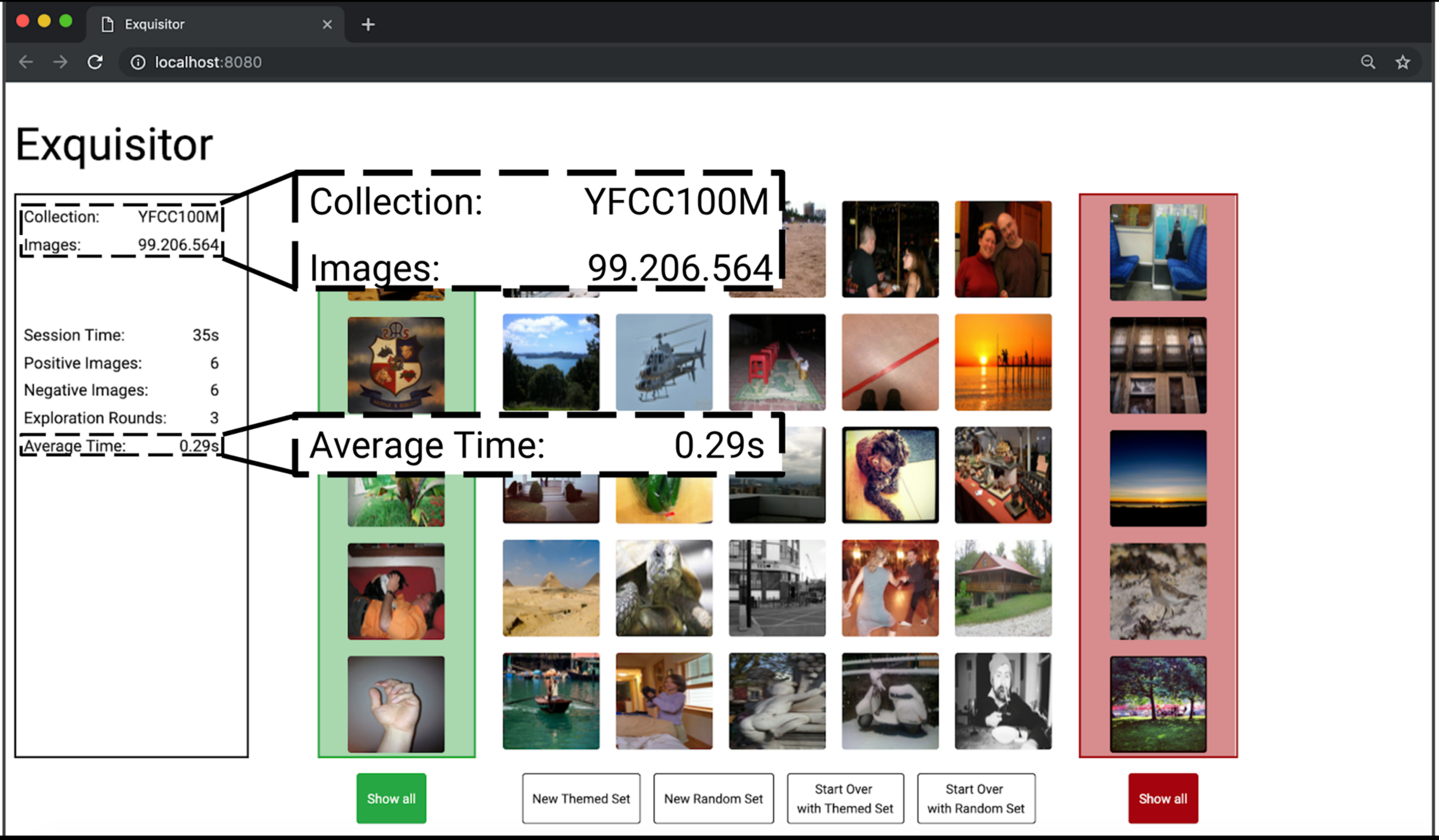}
\vspace{-3ex}
\caption{A screenshot of a prototype interface designed to test the validity of the Exquisitor approach. Exquisitor is capable of interactively learning new analytic categories over the full YFCC100M image collection, with average latency of only 0.29 seconds per interaction, using hardware comparable with standard desktops and modern mobile devices.}
\label{fig:UI}
\end{figure}

At the same time, the way in which people consume  multimedia has been changing drastically. Most multimedia applications today are used on mobile devices with limited computing resources. Nevertheless, the expectation of users is to be able to very efficiently work with their own media collections, as well as with larger external collections. It is therefore imperative that the multimedia community embraces this new trend and provides multimedia tools capable of handling large collections with limited hardware resources.

%\mw{like said in the above, the focus is on mobile devices, but we never explicitly show results on that. So better to change along the following lines: First do a paragraph on difficult to form queries that yield satisfactory results. Then the user relevance feedback. And then comes the catch: users expect the same interactive performance on their standard desktop or mobile devices as for simple query based methods ....... user relevance feedback, At the same time, the way in which people consume multimedia has been changing drastically. Most multimedia applications today are used on mobile devices with limited computing resources. Nevertheless, the expectation of users is to be able to very efficiently work with their own media collections, as well as with larger external collections. It is imperative that the multimedia community embraces this new trend and provides multimedia tools capable of handling large collections with limited hardware resources. }

% \sr{From Marcel: $\downarrow$ If this is the main application I would put this immediately when you start introducing the various applications and large scale. So just after mentioning [59]. You can then mention the general classes of methods similarity search and interactive exploration. Then indicate we do the latter. Later you can then state that for similarity search scalability has been addressed, but only limited for interactive exploration.} 
In this paper, we answer the call for multimedia scalability by addressing the problem of interactive search and exploration in such large collections, using limited hardware resources available to broad audiences.
%A key application area, at such scale, is interactive exploratory analysis of large-scale media collections. 
This is a particularly challenging problem, as with very large collections it is difficult for users to form queries that yield satisfactory results. 
%\jz{$\downarrow$ Minor: I suggest replacing ``originally proposed as'' with ``provides'', I'd argue that relevance feedback didn't come to existence to address the need as described in the previous sentences.}

User relevance feedback, a form of interactive learning, provides an efficient mechanism for addressing various analytic tasks that require alternating between search and exploration. 
%Namely, while in search the user is expected to have a clear information need formulated as a query, exploratio facilitates insight gain inecases where the us en% provides an efficient mechanism for addressing various forms of search as well as the exploratory analysis tasks where the user has no clear information need or is unable or unwilling to formulate it as a query.
% provides a mechanism to solve such 
% exploratory analysis tasks~\cite{638621,huang08} by attempting to understand the information need of the user on the fly and formulate or adjust the query accordingly.
%$\rightarrow$ by attempting to gather and understand the query information from the user and re-formulate his search accordingly.}
%\gtg{I don't like my version better, but I feel clarification of "information need of the user" is in order, somehow.}
Early work on user relevance feedback, however, suffered from both lack of meaningful representations of the media items and lack of high-dimensional indexing techniques for scaling up the feedback loop. 
For example, content based image and video retrieval with user relevance feedback were commonly relying on non-semantic low-level visual features, such as colour, texture, shape and edge histograms \cite{638621}, and using inefficient indexing techniques to facilitate processing, such as R-trees and kd-trees~\cite{410146}. Similarly, even linear models for classification, a popular choice in user relevance feedback applications, were not interactive on collections with, e.g., 100K items~\cite{CC01a}.

%\gtg{ %alternative order for the text below.. 
%There has been relatively little work on user relevance feedback in the last decade, which recently raised serious concerns in the multimedia community \cite{Schoeffmann:2018:IVS:3240508.3241473}. As a result, user relevance feedback using the old methods is only feasible for collections that would be considered small according to today's standards. Recent advances in high-dimensional indexing have however far exceeded these methods in the scale, being applied to collections of up to tens of billions of features, both using a single server and using a computing cluster \cite{GudmundssonMMSys17,7870636,7915742,Jiang:2015:FAC:2733373.2806237}.  Furthermore, advances in data representation, as well as the pressing need for methods to cope with large-scale media collections, clearly imply that the time has come to re-visit the concept of real-time relevance feedback at large scale. % still need to address Marcel's comment on what is the difference between search and exploration. }

%As a result, user relevance feedback was only possible for collections whose scale is \mw{such that with the accuracy of current semantic descriptors are in the realm of fully automatic methods. (just to keep away from considered small today, as of course that is true holds for all datasets from that time interactive or automatic}

There has been relatively little work on user relevance feedback in the last decade, which recently raised serious concerns in the multimedia community \cite{Schoeffmann:2018:IVS:3240508.3241473}. 
However, recent advances in high-dimensional indexing have yielded approaches supporting %that have been applied to 
standard nearest neighbor search in collections with billions of features. 
%, both using a single server and using a computing cluster
\cite{GudmundssonMMSys17,7870636,7915742,Jiang:2015:FAC:2733373.2806237}. 
%\mw{but if applied to 10 billion already nothing is impressive in this paper. So in this paragraph the words interactive  learning should be added} 
Furthermore, advances in data representation, as well as the pressing need for methods to cope with large-scale media collections, clearly imply that the time has come to re-visit interactive learning.

%\mw{one paragraph is needed (also indicated above) to explain the difference between search and exploration and especially that more is needed than just a set of simple queries. } 
In this paper, we present Exquisitor, the first on-the-fly learning approach capable of \textit{interactive exploration of the YFCC100M collection}. 
Overall, our results using the YFCC100M collection show that with the Exquisitor system, the time required to suggest new items in each iteration of the interactive feedback loop is about 4x shorter using 16x fewer computational resources than the state of the art \cite{blackthorn_tmm,jegou:inria-00514462}, a performance improvement of nearly two orders of magnitude, while also improving result quality. Our results thus show that interactive learning is clearly feasible with today's large-scale collections, even on limited hardware. 

In this paper, we make the following major contributions:
\begin{enumerate}
\item Exquisitor significantly surpasses state of the art in interactive search and exploration at large by building on the recent advances in several multimedia areas: data representation from deep learning, data compression from interactive learning, and data retrieval from high-dimensional indexing. 
\item Exquisitor introduces a novel approach to retrieval from cluster-based indexing structures, retrieving the $k$-furthest items from the decision boundary of an interactive classifier.
\item We show, in an experimental evaluation, that 
%using our approach 
suggestions for the interacting user can be produced with sub-second latency using only a single CPU core and very limited memory---hardware resources comparable to today's high-end mobile devices---without any sacrifices in accuracy. 
\end{enumerate}
%\mw{I would make this a paragraph before the contributions as it repeats point 3. If you do it before than point 3 becomes a summary of what has said before in the intro (and that is what is should be)} 

The remainder of this paper is organized as follows. In Section~\ref{sec:sota}, we analyse state-of-the-art methods in interactive learning from a scalability perspective, setting the stage for the Exquisitor approach. In Section~\ref{sec:sys}, we then present the Exquisitor approach in detail, and analyse its performance in Section~\ref{sec:exp}, before concluding.

\section{Related Work}
\label{sec:sota}
%\jz{$\downarrow$ Minor: I suggest ``amateur'' $\rightarrow$ ``casual''. The word amateur here is correctly used, but still has a bit of a negative connotation in colloquial speak I'd say. ``Casual'' sounds more friendly. That said, this is a prime example of a suggestion that can be safely ignored.}
As outlined in the introduction, 
%while great progress has been made in extracting high quality semantic labels from visual content and text at a large scale, 
unlocking the true potential of multimedia collections and providing added value for professional and casual users alike requires joint utilization of interactive learning and high-dimensional indexing. In this section we first describe state of the art in interactive learning.
%and identify potentially productive avenues worth pursuing. 
Then, based on the identified advantages and limitations of interactive learning algorithms, we provide a set of requirements that high-dimensional indexing should satisfy for facilitating interactivity on extremely large collections. Finally, we use those requirements for reflecting on the state of the art in high-dimensional indexing. %In this section we describe state of the art in both fields and provide requirements analysis that motivates our choice of techniques and methodology.

\subsection{Interactive Learning}

Often regarded as an exotic machine learning flavour by the theorists, as it does not fit into the strict ``supervised-unsupervised-reinforcement'' categorization, interactive learning became an essential tool of multimedia researchers from the early days of content-based image and video retrieval \cite{638621,huang08}. While the dominant focus of the research community turned to supervised approaches, which was further encouraged by successes of deep learning, interactive learning nevertheless survived the
% It did not survive the test of time in "something", you do it by "doing something", no? 
% also I use exploratory to distinguish from searching
test of time 
%\gtg{in providing access$\rightarrow$. It did this 
by providing exploratory access to ever-growing multimedia collections as it facilitates e.g. incorporation of human (expert) knowledge \cite{beluch2018power,MIRONICA201638}, the learning of new analytic categories on the fly \cite{blackthorn_tmm,Kovashka2015} and the training of accurate classifiers with minimal number of labeled samples \cite{Yang2015,Wang:2017:CAL:3203306.3203314,beluch2018power}. 
% \sr{references to support this claim. And add something like "as it allows to bring in the human knowledge / subjective xxx / learn categories on which the system has not been trained or something like that.  }
%\jz{$\downarrow$ ``user, which'' $\rightarrow$ ``user, who''}
%\gtg{Could this next paragraph be in the intro? Would that answer Marecel's comment on what is the diff. between large-scale search and exploring?}
Interactive learning involves the process of retrieving items from a collection and showing them to the user, who in turn judges their relevance based on particular criteria, and then using the obtained relevance judgment for modifying or re-training the classifier on the fly. 
The process is repeated as long as the user deems fit for her insight gain.
%The process is repeated until the user is satisfied with the results or the analytic session expires. 
Interactive learning comes in two basic forms, \textit{active learning} and \textit{user relevance feedback} \cite{huang08}.
%\jz{$\uparrow$ I would disagree that the two stopping conditions for a session are user satisfaction or the session expiring. The former is correct, the latter feels like a shortcoming of the system, and it is missing third one, and that one is important: the user gives up, because she feels she's not gaining much by continuing. That's the one that makes interactive learning a challenge IMHO. And there are more options still (the user runs out of time for the exploration, the user has seen the entire collection eventually, but still leaves disappointed in the insight or lack thereof). The shortest way to write this is IMHO ``The process is repeated as long as the user deems fit for her insight gain.''}

\subsubsection{Active Learning}
In active learning, the interacting user annotates samples that will contribute the most to the quality of the final model \cite{Cohn:1996:ALS:1622737.1622744}. In practice, this often means annotating the decision boundary between classes or, in other words, the items for which the classifier is most uncertain \cite{8237827}. The technique, originally proposed in the 90s, recently experienced a revival in the computer vision and multimedia communities as the means of training data-hungry CNNs
%convolutional neural networks 
when obtaining additional labels is costly or unfeasible due to, e.g., a limited time an expert can spend producing annotations \cite{beluch2018power}. The algorithm is commonly trained using a small number of annotated examples and then a fraction of unlabeled items are retrieved and presented to the user for relevance judgment based on a conveniently selected proxy for uncertainty \cite{Yang2015,Wang:2017:CAL:3203306.3203314,beluch2018power}. Despite the effectiveness in training accurate models using small number of samples, active learning is not suitable in our use case of interactive search and exploration. Namely, it does not optimize for the relevance of the items shown to the user in each interaction round (in fact, the opposite), which is one of the main requirements in the design of multimedia analytics systems \cite{Zahalka:2015:AQE:2733373.2806279}.
%\jz{$\uparrow$ ``(on the contrary)'' $\rightarrow$ ``(in fact, quite the opposite)''}

\subsubsection{User Relevance Feedback}
In contrast with active learning, user relevance feedback algorithms present the user, in each interaction round, with the items for which the model (e.g. classifier or regressor) is most confident \cite{638621}. This strategy may require more interaction rounds for the same final quality of the model, but it is more likely to produce relevant items in each interaction round, which is of utmost importance when gaining insight into multimedia collections. Not only is it easier for the user to judge the items for which the model is most confident, but the process of gaining insight is complex, vaguely structured, and incremental, which requires looking at intermediate results rather than final results~\cite{7042476,1626178}.
% \mw{not just for that this is much easier for the user to judge? gaining insight is quite vague and would mean that this is done by looking at intermediate results rather than the final result}
User relevance feedback was frequently 
%seen as 
the weapon of choice in the best performing entries of benchmarks focusing on interactive video search and exploration~\cite{videolympics,8352047}. 
%, such as \cite{videolympics} and, more recently, \cite{8352047}. 
However, those solutions were designed for collections far smaller than YFCC100M, which is the challenge we take in this paper. 

Although attempts have been made to facilitate relevance feedback using CNNs, they are still considered a suboptimal choice for several reasons. Normally they require a large amount of labeled training data, while  users are willing to annotate only a small number of samples in each interaction round. In addition, explainability of results is of utmost importance in analytical tasks, which is why linear models are preferred. Indeed, Linear SVM is still one of the most frequent choices in relevance feedback applications \cite{Kovashka2015,MIRONICA201638,8048559} due to its simplicity and the ability to produce accurate results with few annotated samples and scale to very large collections. 

% To the best of our knowledge, Blackthorn \cite{8048559} is the most efficient
% interactive multimodal learning approach at the moment, capable of
% yielding more accurate results on YFCC100M than, e.g., product quantization \cite{jegou:inria-00514462}---a popular alternative optimized for k-NN search,
% % \sr{more than one ref needed here / or rewrite to call for only one reference}
% while being significantly faster and consuming only modest computational resources. This is achieved through adaptive data compression and feature selection. With this in mind, we conjecture that Blackthorn is the state of the art approach for our use case.

% \sr{From Björn: $\uparrow\downarrow$ Below is a rewrite of the previous paragraph. You can pick which one you like---and then of course edit it\dots}

To the best of our knowledge, Blackthorn \cite{8048559} is the most efficient
interactive multimodal learning approach in the literature. Compared to product quantization~\cite{jegou:inria-00514462}, a popular alternative optimized for k-NN search, Blackthorn was found to yield significantly more accurate results over YFCC100M with similar latency, while consuming only modest computational resources. This performance is achieved through adaptive data compression and feature selection as well as the classification model capable of scoring items directly in the compressed domain. With this in mind, we conjecture that Blackthorn is the state of the art approach for our use case.

%\sr{From Björn: I think 2.2 should in fact be part of 2.1, and made three subsections:
%2.1.1 Active Learning 
%2.1.2 User Relevance Feedback
%2.1.3 Indexing Requirements}

%\subsection{Requirements for High-Dimensional Indexing}
\subsubsection{Requirements for Indexing}
\label{sec:sota:ix}

%\jz{$\downarrow$ I have tried to tie this into the text better, please see below and review (I am not a high-dim indexing guy, might have missed something).}
%\mw{as indicated I feel that this whole section should be related work. This means that indexing requirements should be introduced as a prelude to the high dimensional indexing subsection. Question is why there are no requirements on the interactive learning part. }
The most computationally intensive part of the interactive learning process is the selection of candidates to show to the user. This process must in principle examine the feature vectors of all items of the media collection, while eventually only a tiny fraction of the large collection is shown to the user. Thus, there remains a potential for performance improvement that interactive learning on its own does not tap into: utilizing the inherent structure of the feature vector collections.

%\gtg{
%Efficient utilization of such inherent data structure is the domain of high-dimensional indexing. Some early relevance feedback approaches indeed considered indexing feature vectors for efficient scoring. Since indexing approaches have generally only supported similarity-based queries, however, index-based relevance feedback approaches typically focused on refining queries rather than classification boundaries.
%$\xrightarrow{}$
Efficient utilization of the inherent structure of data is the domain of high-dimensional indexing and some early relevance feedback approaches did indeed consider indexing feature vectors for efficient scoring. However, since indexing approaches at that time generally only supported similarity-based queries, index-based relevance feedback approaches typically focused on refining search queries rather than  classification boundaries.
%} % I'm lacking the word here for what the XXX dows with the classification boundaries but I read the old version again and again and there is just something off, I was not getting-it :/
Furthermore, due to lack of scalable high-dimensional indexing methods in the past, these approaches were always limited to small collections only.  

As we look towards today's scalable high-dimensional indexing approaches as a potential source of performance improvements, we have identified the following requirements for a successful high-dimensional indexing approach that enhances the performance of interactive learning:

%\jz{(wearing an evil-reviewer's hat): I'd say that R1, written as it is, is dubious. I personally see its merit (and wearing the software dev hat, I wholeheartedly agree), but the argument that response time guarantees are more valuable than result quality guarantees is very strong, but not substantiated enough. It is only supported by a workshop paper. Written like this, as the nasty reviewer I could say ``riiight, so it's another example of some guys introducing a new viewpoint and then bashing all the related work because they haven't explicitly considered this before even though this is only some workshop paper''.}
%\jz{(cont'd for readability) My suggestion: keep R1 as it is, just change the last sentence to sth like ``A successful approach in the interactive learning setting combines good result quality with response time guarantees [60].'' Then, R1 becomes indisputable and it still works 100\% in the text following it.}

\begin{itemize}
\item [\reqsrt] \textit{Short and Stable Response Time:~}  The highly interactive nature of the process demands not only a short response time, but also \textit{predictable} response time, to avoid distracting users.\footnote{The requirement for predictable latency is well known from other domains~\cite{DBLP:conf/fast/BjorlingGB17,Barroso:2017:AKM:3069398.3015146,DBLP:conf/sisap/AmsalegJL18}.} The approximate nature of the queries and features, on the other hand, limits the impact of result quality guarantees in the high-dimensional space. A successful approach in the interactive learning setting combines good result quality with response time guarantees~\cite{tavenard:inria-00576886}.
%Response time guarantees are thus far more valuable than result quality guarantees~\cite{tavenard:inria-00576886,DBLP:conf/fast/BjorlingGB17,Barroso:2017:AKM:3069398.3015146,DBLP:conf/sisap/AmsalegJL18}.

\item[\reqpqe] \textit{Preservation of Feature Space Similarity Structure:~} The purpose of interactive classifiers is to capture the user intent as it evolves during the interactive session. 
The classifiers capture this intent using a hyperplane that attempts to separate relevant items identified in previous interaction rounds from the rest of the collection, and compute relevance of the remaining items based on the similarity structure of the feature space. The space partitioning of the high-dimensional indexing algorithm must preserve this similarity structure.

\item [\reqkfn] \textit{$k$ Farthest Neighbours:~} 
As discussed in the above, relevance feedback approaches typically request the items farthest from the classification boundary. Furthermore, as the results are intended for display on screen, the index must return exactly $k$ farthest neighbours ($k$-FN). Finally, since the interactive classifier is an approximation of the analyst's intent, approximate answers are also acceptable.
\end{itemize}

We are not aware of any work in the high-dimensional literature specifically targeting approximate $k$-FN where the query is a classification boundary. We therefore next review the related work and discuss how well different classes of high-dimensional indexing methods can potentially satisfy these three requirements.

\subsection{High-Dimensional Indexing}
\label{sec:ix}

Due to the curse of dimensionality, scalable high-dimensional indexing methods must rely on approximate  similarity searches, typically trading off small reductions in quality (or even just quality guarantees) for dramatic response time improvements. In this section we therefore only consider approximate methods.

\subsubsection{Approximate Nearest-Neighbour Queries}
Most high-dim\-ensional indexing methods rely on some form of quantization. The first group of methods uses scalar quantization. LSH, for example, uses random projections acting as locality preserving hashing functions~\cite{Gionis:1999:SSH:645925.671516, Datar:2004:LHS:997817.997857,10.1109/FOCS.2006.49,SDI06}. Its performance mainly depends on the quality and the number of hashing functions in use. Hence, many approaches improve hashing~\cite{DBLP:conf/nips/WeissTF08,DBLP:conf/cvpr/JainKG08,Tao:2009:QEH:1559845.1559905,DBLP:conf/cvpr/WangKC10,DBLP:journals/prl/PauleveJA10,5767837,DBLP:conf/iccv/JinHLZLCL13}, whereas others reduce the number of hash functions~\cite{Lv:2007:MLE:1325851.1325958,Joly:2008:PML:1459359.1459388}. 
LSH and similar methods, however, fail to satisfy the three requirements: they focus on quality guarantees rather than performance guarantees (\reqsrt); hashing creates ``slices'' in high-dimensional space, making ranking based on distance to a decision boundary impossible (\reqpqe); and they typically focus on $\epsilon$-range queries, giving no guarantees on the number of results returned (\reqkfn). While too many answers can be handled by filtering, too few answers may also be returned.

The NV-tree is another high-dimensional indexing method which also uses random projections at its core~\cite{10.1109/TPAMI.2008.130,Lejsek:2011:NNN:1991996.1992050}. It recursively projects points onto segmented random lines and stores the resulting buckets onto disk. The NV-tree is a disk based method, designed for  collections larger than RAM, and has been shown to outperm LSH for large-scale indexing~\cite{10.1109/TPAMI.2008.130}. The NV-tree satisfies \reqsrt{} and \reqkfn{} well, but its leaves have irregular shapes and do not satisfy \reqpqe.

A second group of methods is based on vector quantization, typically using clustering approaches, such as $k$-means, to determine a set of representative feature vectors to use for the quantization. 
%\mw{better: a typical clustering method used for that purpose is xxx? as k-means is not vector quantization}
These methods create Vorono{\"\i} cells in the high-dimensional space, which satisfy \reqpqe{} very well.  Some methods, such as BoW-based methods, only store image identifiers in the clusters, thus failing to support \reqkfn, while others satisfy that requirement by storing the features and ranking the results in the nearest (or farthest) clusters. Finally, many clustering methods seek to match well the distribution of data in the high-dimensional space. Typically, these methods end with a large portion of the collection, often more than 20\%, in a single cluster, which in turn takes very long to read and score, thus failing to satisfy \reqsrt. The extended Cluster Pruning (eCP) algorithm, however, is an example of a vector quantifier which attempts to balance cluster sizes for improved performance, thus aiming to satisfy all three requirements. 

Product quantization (PQ)~\cite{jegou:inria-00514462} and its many variants~\cite{DBLP:journals/tmm/XioufisPKTV14,DBLP:journals/pami/GeHK014,DBLP:conf/cvpr/KalantidisA14,DBLP:conf/cvpr/HeoLY14, DBLP:conf/cvpr/BabenkoL12,DBLP:journals/pami/BabenkoL15} cluster the high-dimensional vectors into low-dimensional subspaces that are indexed independently.  Compared to hashing based methods, the ones relying on clustering better capture the location of points in the high-dimensional space, which in turn improves the quality of the approximate results that are returned. One of the main aims of PQ is compression of the data, however, and PQ-based methods 
%neither guarantee the number of items returned (\reqkfn) nor the time taken to retrieve them (\reqsrt). Furthermore, these methods 
essentially transform the Euclidean space, complicating the identification of furthest neighbours (\reqpqe).  
%\gtg{In~\cite{blackthorn_tmm}, product quantization was compared directly with Blackthorn's compression method; the results showed that with similar compression levels, product quantization yielded significantly inferior results. Product quantization thus fails to satisfy the three requirements above. $\rightarrow$
In~\cite{blackthorn_tmm}, PQ-compression was compared directly with the novel compression method proposed for Blackthorn; the results showed that with similar compression levels, PQ-compression yielded significantly inferior result quality. As PQ-compression is a pre-requisite for using PQ, it does not appear to be a promising candidate for user relevance feedback.
%BT was introduced way way to casually here, it has never been mentioned by name before this point. 
%\jz{$\uparrow$ I'd say PQ does satisfy R3. It is a $k$-NN approach, and from my experience you can guarantee exactly $k$ neighbours easily. Also, PQ not satisfying R1 is dubious - yes, the time per query might vary slightly, but they are all safely sub-second even on larger collection, so if you guarantee let's say 1 second as a response time, you can do R1 with PQ (and just wait the remainder of the time after the query). What we say here is true about ``interactive learning using PQ-compressed features,'' but I'd say this is not evident from the text that we're talking about this and not PQ as a whole.}

\subsubsection{Hyperplane-Based Nearest-Neighbour Queries}
Some researchers have considered this problem: given a collection of high-dimensional points, which are closest to a hyperplane cutting through the high-dimensional space? This problem is central to various \textit{active learning} tasks, where the goal is to request labels for those points that are most informative, as described above, %maximizing the profitability of the next pooling 
which in turn helps find the most appropriate SVM decision boundary. Projections and hash functions have been proposed~\cite{DBLP:journals/imst/CrucianuEOT08, DBLP:conf/nips/JainVG10, DBLP:journals/pami/BasriHZ11,DBLP:journals/pami/Vijayanarasimhan0G14}, which means that these hyperplane-based approaches from the literature are not applicable to user relevance feedback, based on the analysis above.

\subsubsection{Farthest-Neighbour Queries}
Farthest neighbour queries are in part motivated by the need to improve the diversity of what is returned to users, e.g., in applications making use of collaborative filtering for product recommendation~\cite{DBLP:conf/www/AbbarAIM13, DBLP:conf/compgeom/AbbarAIMV13, DBLP:conf/cscw/SaidFJA13}. The farthest neighbours problem consists of finding the vectors from a data set that maximize the distance to a query point. Approximate solutions~\cite{DBLP:journals/comgeo/AgarwalMS91, DBLP:conf/soda/Indyk03, DBLP:conf/sisap/PaghSSS15, DBLP:conf/sisap/CurtinG16, DBLP:journals/jcst/XuBYZTGX17,DBLP:journals/is/PaghSSS17}, based on hashing or exploiting the distribution of the data are often prefered to exact ones~\cite{DBLP:conf/icalp/Williams04, DBLP:books/lib/BergCKO08}, which are extremely costly to compute. Some methods are named \emph{c-Approximate} as they return vectors that are at least $1/c$ times the distance of the query point to its true furthest neighbour~\cite{DBLP:conf/soda/Indyk03,DBLP:journals/is/PaghSSS17}. As before, these methods fail to satisfy the three requirements.  

\subsection{Summary}
Based on the requirements above, and our analysis of the state of the art in high-dimensional indexing, we believe that cluster-based approaches, such as eCP, are the best candidates for relevance feedback. These approaches, however, have never before been used for $k$-farthest neighbour queries from a decision boundary.% in the literature.

\newlength{\figwid}
\setlength{\figwid}{\columnwidth}

\begin{figure}
	\includegraphics[width=\columnwidth]{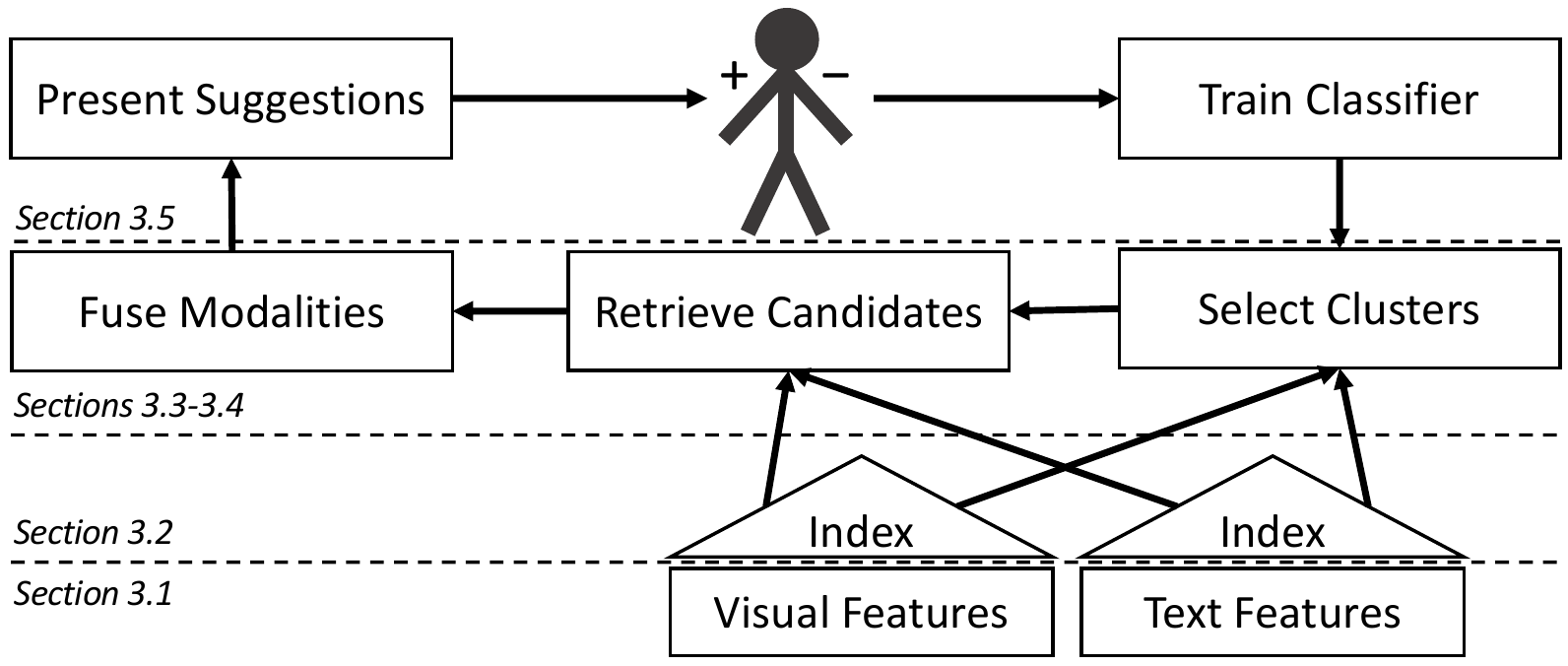}
	\caption{Overview of the Exquisitor approach.}
	\label{fig:exq}
\end{figure}

\section{The Exquisitor Approach}
\label{sec:sys}
% \mw{I miss a figure of the overall approach. Make it such that all subsections have a place.}
%\mw{you need some more equations / definitions in the text below. All now is descriptive (this is what we engineered) but doesn't highlight the method that is actually used. To reproduce the work in this paper I would have to go through all the underlying papers. In addition that pseudocode would help.}

In this section, we describe Exquisitor, the first active learning approach in the literature capable of interactive learning over the YFCC100M image collection using hardware resources similar to those found in high-end mobile devices.\footnote{While many current mobile devices have only 2--4GB of RAM, the trend is clearly towards larger RAM: the latest Samsung Galaxy models are equipped with 8GB of RAM, while the upcoming models will have 12GB~\cite{forbes}.}
%\gtg{ You must clarify what you mean here.. What mobile platform are we talking about? Examples: iPhone 7 has 2GiB, X and 8 have 3GiB and the latest, XS has 4GiB. How much RAM would we need?}
Figure~\ref{fig:exq} shows an outline of the Exquisitor approach, and also an outline of this section. We start by considering the multimodal data representation, then describe the indexing and retrieval algorithms, before describing the choice of suggestions and interactions with the user. 

To facilitate the exposition in this section, we use actual parameters and settings from the YFCC100M collection in various places, as this allows us to discuss several practical issues that arise when dealing with such a large and unstructured image collection. Needless to say, however, the Exquisitor approach can handle any image collection, including much larger collections than YFCC100M. 

\subsection{Image Representation}
The YFCC100M collection contains 99,206,564 Flickr images, their associated annotations (i.e. title, tags and description), and a range of metadata produced by %either 
the capturing device, the online platform, and the user (e.g., geo-location and time stamps). Following recent literature,
%state of the art, 
each image is represented by two semantic feature vectors. The visual content is encoded using 1,000 ILSVRC concepts~\cite{Russakovsky2015} extracted using the GoogLeNet convolutional neural network~\cite{7298594}. The textual content is encoded by a) treating the title, tags, and description as a single text document, and b) extracting 100 LDA topics for each image using the gensim toolkit \cite{rehurek_lrec}. 

%\jz{$\downarrow$ I understand the need of the shortening, but ``880GB of RAM ... beyond the storage capacity'' reads weird, as you do not store RAM (yes, we can discuss this endlessly, ``what about dumps/swap,'' but you know what I mean). How about ``Directly working with these representations, however, is infeasible. They require around 880GB of memory, which not only hardly fits in RAM, but also is well beyond the storage capacity of small devices.''}
Directly working with these representations, however, is infeasible. They require around 880GB of memory, which not only hardly fits in RAM, but also is way beyond the storage capacity of mobile devices. The feature vectors are therefore compressed using the methodology presented in~\cite{8048559}. By storing only the 6 most important features of each feature vector, along with the feature identifiers, and by using a compression method based on the ratio between feature values, each feature vector can be represented using only three 64-bit integers, resulting in 48 bytes per image, or about 4.8GB in total.
%\gtg{FYI: 4.8 GiB is more than available RAM on any Apple mobile phone to date.} 

%\jz{$\downarrow$ This needs another [73] (Blackthorn) IMHO. It is a separate paragraph which is taken pretty much directly from the TMM article, and it only has an indirect (albeit logical) ref from the previous one. Careful about plagiarism rules here.}
\newcommand{\tfidf}{\mbox{\textit{tfidf}}}
The most important features are selected using a TF-IDF based approach, where the strongest features appear with high confidence in a few images~\cite{8048559}. Consider a feature $f$ with average value $\mu_f$ and standard deviation $\sigma_f$ across all the items in the collection. For an individual item $X$, the TF- IDF score of that feature is given by Equation~\ref{eq:ifidf}, where $N$ is the number of items in the collection $C$ and  $\llbracket\cdot\rrbracket$ is the Iverson bracket:
\begin{equation}
\tfidf(x_f) = x_f \log\left(1+ \frac{N}{\sum_{Y\in C}\llbracket y_f > \mu_f + \sigma_f \rrbracket}\right) 
\label{eq:ifidf}
\end{equation}
The TF portion is thus the value of the feature itself, while the IDF portion determines the feature’s rarity, represented by the fraction of the collection where the feature is strongly present. The features of each media item are sorted by $\tfidf(x_f)$ in descending order and the top 6 features are selected to represent the item. 

%\jz{$\downarrow$ This is a nice paragraph. Introduces novelty (nothing like this is in the Blackthorn paper) and shows mastery of the huge dataset, making the reader more confident that we know what we are talking about.}
The two compressed feature vector collections have some interesting properties worth mentioning, that would not occur in a smaller collection that was easier to curate. First, some proportion of the images in the collection have been removed from Flickr, and therefore are represented in the collection using a standard ``not found'' image.\footnote{The image collection was actually downloaded very shortly after release, but already then this had become a significant issue.  Interestingly, the classification concept most strongly associated with the ``not found'' images is ``menu''.} As a result, the visual feature vectors for these missing images are identical and if one is considered a candidate in the visual domain, they all are, potentially crowding out more suitable candidates. Second, a similar situation arises in the textual domain, where many images have no text tags, and hence their textual feature vector is all zeros. Third, due to the lower dimensionality of the textual feature vectors, the likelihood of two images having the same textual feature vector is much higher than the likelihood of two images having the same visual feature vector. As we show below, all these properties impact the cluster size distribution significantly, which in turn impacts the time required to propose suggestions. In short, the more even the distribution, the less processing time is required. However, our results show that even with these properties, very short latency is achieved.
%the time required to propose suggestions is very small.

\subsection{Data Indexing}

The data indexing algorithm used in Exquisitor is based on the extended Cluster Pruning (eCP) algorithm~\cite{Gudmundsson:2010:LPS:1878137.1878145,4587635}. As motivated in Section~\ref{sec:sota}, the goal is to individually cluster each of the two feature collections with a vectorial quantizer, using a hierarchical index structure to facilitate efficient selection of clusters to process for suggestions. The clusters for each collection are formed by randomly picking a set of feature vectors, called {\em representatives}, from the collection $C$, and then assigning all feature vectors to these representatives based on proximity. The Euclidean distance function has been implemented directly in  compressed space and used as the discriminative distance function for eCP.
%\jz{$\uparrow$ The last sentence feels like a gloss-over. Adapted how? (I know, but the reader might not) Either expand on that (maybe in a journal extension :)), or provide a two-sentence description. Also, there is a high possibility that there was a sparse-Euclidean dst used somewhere before, adding a ref and saying like ``inspired by [REF], we have adapted the Euclidean dst function to operate on the compressed data repre''}

To facilitate the assignment to clusters---as well as the subsequent retrieval from clusters---an index is created using Algorithm~\ref{alg:ci}. When calling CreateIndex($C\,$), the algorithm recursively selects 1\% of the features at each level as representatives for the level above, until fewer than 100 representatives remain to form the root of the index. The bottom level of the index for each modality in the YFCC100M collection thus consists of $992,066$ clusters, organized in a $3$ level deep index hierarchy, which gives on average $100$ feature vectors per cluster and per internal node.

%\jz{$\downarrow$ I like this argumentation. Also, to further support the choice, I would maybe remind the reader again of the scale, because him coming to this page with the ``trivially-looking'' (not my opinion, putting myself in the reviewer's shoes) algorithm could lead in this ``huh? that's it? easy no novelty $\rightarrow$ reject.'' A statement saying that lengthy, detailed computations are intractable on 100M could go a long way. But this is an ignorable suggestion.}
Two notes are in order. First, when building the indices, the average cluster size was chosen to be small, as previous studies show that searching more small clusters yields better results than searching fewer large clusters~\cite{Gudmundsson:2012:IST:2261008.2261018,EMMA}. Second, eCP is essentially the first step of the $k$-means algorithm.
%, although the $k$ in $k$-means is typically much lower and therefore an index is not needed. 
The reason for avoiding the refinement iterations of $k$-means---in addition to efficiency---is that the cluster size distribution tends to become more skewed as more iterations are completed, and multiple works from the literature have shown that skewed cluster size distributions are anathema to \textit{stable} response time~\cite{Gudmundsson:2010:LPS:1878137.1878145,tavenard:inria-00576886,EMMA,DBLP:conf/eccv/BaranchukBM18,DBLP:conf/sisap/AmsalegJL18}. This is particularly important in the YFCC100M setting, as the feature collections already have some inherent skew, as mentioned above, which the indexing approach should not further aggravate. %We discuss handling the inherent skew in more detail below.

\begin{algorithm}[t!]
	\DontPrintSemicolon
%	\KwData{Collection $C$}
%	\KwResult{Index $I$}
    \lIf{$|C| < 100$}{Return new root node with $C$}
	\Else{
	    $C’ \leftarrow$ = 1\% sample of $C$\;
	    $I \leftarrow$ CreateIndex($C’$)\;
	    \lFor{$X’ \in C’$}{Create new sub-node for $X’$ in $I$}
	    \lFor{$X \in C$}{Add $X$ to the nearest sub-node in $I$}
	    Return $I$}
	\caption{CreateIndex(Collection $C\,$) $\rightarrow$ Index $I$}
	\label{alg:ci}
\end{algorithm}

\subsection{Suggestion Retrieval}
The retrieval of suggestions has the following three phases. First, the most relevant $b$ clusters are identified, then the most relevant $r$ candidates for each modality are identified, and finally the most relevant $k$ suggestions are selected from the set of candidates using modality fusion. Each phase is described below; the next subsection then discusses some extension to the basic retrieval method, including how to handle the extreme skew in the cluster size distribution. As noted in Section~\ref{sec:sota}, to the best of our knowledge this is the first instance of using cluster-based indexing approaches to facilitate $k$-farthest neighbors from a hyperplane.
%Note that the high-dimensional index allows for efficient dynamic loading of the clusters into memory, enabling cluster scoring in interactive time. 

\subsubsection{Identify $b$ Most Relevant Clusters:} 
The number of top clusters considered can be adjusted by a search expansion parameter $b$, which affects the size of the subset that will be scored. This parameter can be used to balance between search quality and latency at run-time. In each iteration of the interactive learning process, the index of representatives is used to identify, for each modality, the $b$ clusters most likely to contain useful candidates for suggestions. 
% Motivated by the requirements analysis from Section~\cite{sec:sota} and consistent with the state of the art in large-scale user relevance feedback, 
As described in Section~\ref{sec:sys:api}, the classifier used in Exquisitor is Linear SVM; 
the dot-product computations to score representatives (and then feature vectors) are done directly in the compressed space and the $b$ clusters farthest from the separating plane (in the positive direction) are selected as the most relevant clusters.

\subsubsection{Select $r$ Most Relevant Candidates per Modality:}
Once the most relevant $b$ clusters have been identified, the compressed feature vectors within these clusters are scored to suggest the $r$ most relevant media items for each modality. The method of scoring individual feature vectors is the same as when selecting the most relevant clusters; an unordered list of the $r$ most relevant items is dynamically maintained throughout the scoring process. 

Some notes are in order here. First, in this scoring phase, media items seen in previous rounds are not considered to be candidates for suggestions. Second, an item already seen in the first modality is not considered as a suggestion in the second modality, as it has already been identified as a candidate. Third, if all $b$ clusters are small, the system may not be able to identify $r$ candidates, in which case it simply returns all the candidates found in the $b$ clusters.

\subsubsection{Modality Fusion for $k$ Most Relevant Suggestions:} 
Once the  $r$ most relevant candidates from each modality have been identified, the modalities must be fused by aggregating the candidate lists to produce the final list of $k$ suggestions. First, for each candidate in one modality, the score in the other modality is computed if necessary, by directly accessing the compressed feature vector, resulting in $2r$ candidates with scores in both modalities.\footnote{To facilitate late modality fusion, the location of each feature vector in each cluster index is also stored as an array; each such vector requires about 800KB of RAM.} Second, the rank of each item in each modality is computed by sorting the $2r$ candidates by the score in the modality. Finally, the average rank is used to produce the final list of $k$ most relevant suggestions, thus favoring items that score relatively well in both~modalities.

\subsection{Retrieval Extensions}
\label{sec:sys:ext}

We now describe three improvements to the basic retrieval algorithms, which aim at improving both latency and result quality, by addressing practical issues that arise in this real-life setting.

\subsubsection{Multi-Core Processing:}
Exquisitor can take advantage of the availability of multiple CPU cores.
%, when that opportunity arises. 
With $w$ cores available, the system creates $w$ workers and assigns $b/w$ clusters to each worker. Each worker produces $r$ suggestions in each modality and fuses the two modalities into $k$ candidates, as described above. The top $k$ candidates overall are then selected by repeating the modality fusion process using all $w\cdot k$ suggestions from the $w$ workers. 
%Our results show that the Exquisitor system is so efficient that only limited benefits are seen for the YFCC100M collection, but preliminary experiments with a collection of 1 billion items shows that multi-core processing impacts performance at that scale.

%\jz{$\downarrow$ Again, nice, particular analysis, but add YFCC100M somewhere to the intro of this subsection. Otherwise, it might seem as if we are giving general statements.}
\subsubsection{Handling Skew:} 
\begin{figure}
\includegraphics[width=.9\columnwidth]{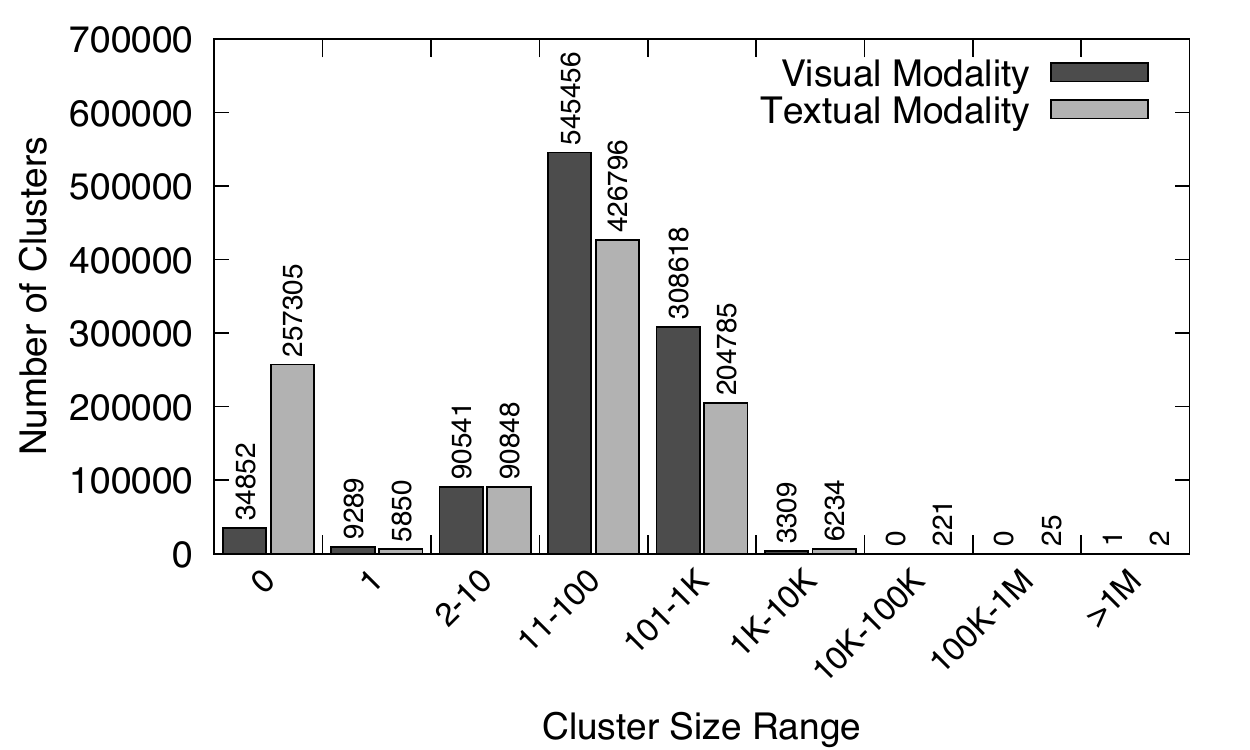}
\caption{Distribution of cluster sizes in the two domains.}
\label{fig:e2:skew}
\end{figure}

As described above, with the YFCC100M collection, both modalities have 1-2 clusters that are very large, with more than 1M items. These clusters require significant effort to process, while contributing negligibly to the quality of results.  Furthermore, in the text domain, many images may have identical feature vectors, resulting in high variability of the cluster sizes.
To quantify the amount of such data \textit{skew}, Figure~\ref{fig:e2:skew} shows the number of clusters in each size range for each modality. Recall that both cluster indexes are created such that the average cluster should contain 100 feature vectors. Consider first the visual modality. As already mentioned, one cluster of ``not found'' images contains over 3M feature vectors. The second largest cluster, however, contains less than 10K feature vectors, and more than 85\% of all clusters range from 11 to 1000 feature vectors. Note that about 35K clusters have 0 feature vectors. The representatives of these clusters are most likely all found in the large 3M+ cluster; such empty clusters are always omitted from consideration. 

Turning to the the text modality, Figure~\ref{fig:e2:skew} shows that the cluster size distribution is significantly more varied, with more large clusters and more empty clusters, but fewer clusters of mid-range sizes. As the empty clusters are ignored, more feature vectors are processed, on average, for the text modality. In the final processing of suggestions, however, both modalities are weighted equally.

With a single worker, the best strategy for handling large clusters is to avoid processing them. In Section~\ref{sec:exp:e2}, we explore the impact of omitting clusters above a size threshold $S_m$ on quality and performance. 
With multiple workers, we can also explore applying multiple workers to the largest clusters, while assigning small clusters to the remaining workers; this is future work.
%we leave this exercise for future work.

\subsubsection{Improving Quality:}
We observe that media items that score moderately highly in both modalities (in the following, we refer to these as \textit{bi-modal} media items) are more likely to be relevant than media items with a high score on one modality, but a low score on the other (we refer to these as \textit{mono-modal} media items).  In a setting where all $b$ clusters are processed by one worker, there is a risk that 
%of missing the bi-modal media items, as 
the candidate lists in each modality may be dominated by  mono-modal items. When the mono-modal items happen to be part of a large cluster, the likelihood that bi-modal items are suggested becomes quite low. Note that turning to a median rank aggregation scheme based on the Condorcet criterion is not a solution in this case, as finding the bi-modal items would require reading a very substantial portions of the collection~\cite{DBLP:conf/mm/LejsekAJA06a}.

In a multi-worker setting, however, each worker produces its own set of suggestions, which are subsequently merged to produce the final result. The first $b/w$ clusters thus result in one set of candidates, the next $b/w$ clusters result in another set, and so forth. This means that the first 1-2 workers are likely to produce candidates that are mono-modal, while the remaining workers, which are processing moderately relevant clusters, are more likely to suggest bi-modal items. Indeed, our results show that in the basic configuration where each worker produces only one set of candidates, the result quality improves as workers are added, and the results with 16 workers are significantly better than with 1 worker.

We address this problem by introducing a new parameter $S_c$ to segment the clusters to be processed. Each worker then processes $S_c/w$ such segments. When a single worker is used ($w=1$), that single worker then produces $S_c$ rounds of suggestions, each from $b/S_c$ clusters. With this approach, result quality is independent of the number of workers.
%\jz{$\uparrow$ I think we are skipping ahead with the last sentence, this should be a post-experiment statement. I'd say sth like ``With this approach, the results are further diversified and independent of the number of workers.''}

\subsection{Relevance Judgment and Learning}
%\mw{API doesn't seem to be part of this section}
\label{sec:sys:api}
%\jz{If it's needed to save space, this can be easily cut and/or partly merged with previous text. I think we're repeating related work at this point pretty much.}
%Motivated by the requirements analysis from Section~\ref{sec:sota} and 
Consistent with the state of the art in large-scale user relevance feedback, the classifier used in Exquisitor is linear SVM.
%The underlying interactive learning algorithm is a linear SVM. 
The choice is further motivated by the algorithm's speed, reasonable performance and compatibility with the sparse compressed representation. 
%Exquisitor employs interactive multimodal learning to provide suggested items to the user that are as relevant to her as possible. The interaction setting corresponds to standard relevance feedback: 
In each interaction round, the user is provided with a set of suggested items, marks the relevant and not relevant ones and submits those labels to the system. The system then takes the user's 
%``relevant'' and ``not relevant'' 
labels, uses them as positive and negative training examples (the set of negative examples can also be augmented with a random selection from the large collection), trains an interactive classifier, and provides a new set of suggested items, avoiding
%based on it whilst not repeating 
items seen in previous rounds. 

\section{Experimental Evaluation}
\label{sec:exp}

In this section, we experimentally analyse the performance and quality of 
Exquisitor
%, and associated parameters described in the previous section, 
with the YFCC100M collection. We first outline the experimental protocol followed, before describing the results of three key experiments, seeking to answer the following questions:
%\mw{in this list comparison to the state-of-the-art is missing}
\begin{enumerate}
\item How does the performance of the Exquisitor approach compare to the state of the art in interactive learning, and what is the influence of the number of clusters ($b$)  on the tradeoff between latency and result quality?
\item What is the impact of addressing skew, by omitting large clusters from processing, on  latency and result quality? % of Exquisitor?
\item What is the impact of applying additional CPU cores on the latency of Exquisitor? 
\end{enumerate}
%\mw{I always find it very strange to already use the results before you have them. Also in line with what Honza is saying just explain the obvious tradeoff here.}
%As the results show, the Exquisitor system has both better interactive performance and result quality than the state of the art approach from the literature~\cite{blackthorn_tmm}.  The results also show, however, that there is a tradeoff between the interactive performance and result quality, where some parameter settings give a significant improvement in the interactive performance, at a cost of some reduction in quality.
%\jz{$\uparrow$ Very minor: The ```the results also show, however'' sentence is written as if it is a big surprise that there is a trade off between performance and accuracy, whilst I would say this is expected. This is not a negative result, I would spin it as Exquisitor having an advantage because it is further customizable on this tradeoff through parameters (which are well defined).}
%\la{$\uparrow$ This paragraph could be removed.}

\begin{figure}
	\includegraphics[width=\figwid]{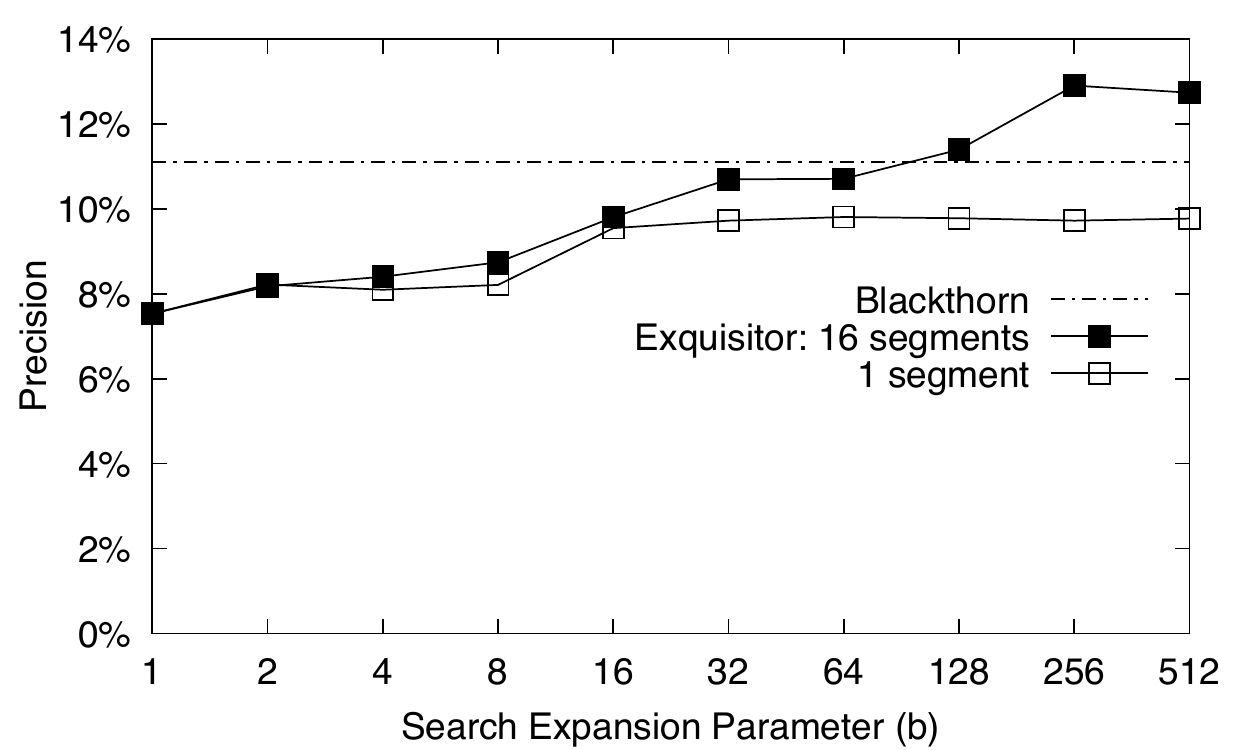}
	\caption{Average precision over  first 10 rounds of analysis.\hfill \\(Experiment 1: YFCC100M; varying $b$; $w=1$; $S_c=1, 16$).}
	\label{fig:e1:p}
\end{figure}

\subsection{Experimental Setup}
%As motivated in the introduction, the YFCC100M is by far the largest publicly available collection of multimedia items \cite{Thomee:2016:YND:2886013.2812802}. The YFCC100M collection contains approximately 99.2 million Flickr images, their associated annotations (i.e. title, tags and description), and a range of metadata produced by either the capturing device, the online platform, or the user (e.g., geo-location and time stamps). From each images we have extracted 1000 ILSVRC concepts \cite{Russakovsky2015} using the GoogLeNet convolutional neural network \cite{7298594}. Treating the title, tags, and description as a single text document, we have then extracted 100 LDA topics for each image using the gensim toolkit \cite{rehurek_lrec}. To create the compressed representation, we then selected the top 6 visual concepts and LDA topics with the highest detection score, according to the tf-idf methodology presented in the previous section.  
%The index for each modality consists of $992,066$ clusters, organized in a modest $3$ level deep index hierarchy, which gives on average $100$ feature vectors per cluster for the YFCC100M dataset.

%\jz{$\downarrow$ Even though I toned down the first sentence with ``to the best of our knowledge'', it still seems like a strong, risky statement to me. Might be worth checking recent work on YFCC100M, as \textit{any} experimental work (even vaguely- or non-related to ours) invalidates this statement. Can't do it myself I'm afraid - don't have university access to papers anymore.}
As the literature, to the best of our knowledge, contains only one experimental interactive learning protocol that has been applied to the YFCC100M collection~\cite{blackthorn_tmm}, we have chosen to follow that protocol. This evaluation protocol is inspired by the well-known MediaEval Placing Task \cite{Larson:2011:ATG:1991996.1992047,451edeadebaa4360865c0527aeafb5c9}. 
%We note that although this approach is not designed for estimating capturing location of the images, which is the focus of the Placing Task, we find it suitable  for general evaluation of analytic performance, and it allows a direct comparison with the results from~\cite{blackthorn_tmm}. 
From the 2016 edition of the MediaEval Placing Task, we have created artificial \textit{actors} (or users) by selecting the 50 world cities represented with the largest number of images in the YFCC100M collection. The \textit{relevance set} of each artificial actor then consists of the images and their associated metadata captured within 1000km from the centre of one city, which is the largest radius used in the Placing Task. A large radius was intentionally selected due to our focus on semantic relevance of the items instead of their exact capturing location.
%, as nearby cities tend to be similar. 
%\mw{That still feels like a weird motivation. Paris and London are similar to Amsterdam? \bj{Fair enough. Not much we can do at this point.}}
For each actor, the evaluation starts by pre-training the interactive classification model using 100 randomly selected relevant images as the positives and another 200 negative examples randomly selected from the collection. In each interaction round the actor is then presented with the 25 items considered most relevant by the model. As interface design is not the focus of this paper, we choose this number with, e.g., a basic 5x5 grid visualization in mind. Then, the items that are part of the actor's profile are added to the set of positives and 100 randomly selected items are used as  negatives to train the interactive learning model in the subsequent round. 

To illustrate the tradeoffs between the interactive performance and result quality, we focus our analysis on precision and latency (response time) per interaction round.  It is worth noting that due to both the scale of YFCC100M and its unstructured nature, precision is 
%generally 
lower than in experiments involving small and well-curated collections. The important comparison is therefore to Blackthorn~\cite{blackthorn_tmm}, the only state of the art algorithm capable of handling YFCC100M with interactive performance. 

Both  Exquisitor  and  Blackthorn  are compiled with g++. All experiments are performed using dual 8-core 2.4 GHz CPUs, with 64GB RAM and 4TB local SSD storage. Note, however, that the collection used in the experiments requires less than 7GB of SSD storage and RAM, and in most experiments  Exquisitor  uses only a single CPU core.
%; these  computing resources are similar to that of today's mobile devices.

\begin{figure}
\includegraphics[width=\figwid]{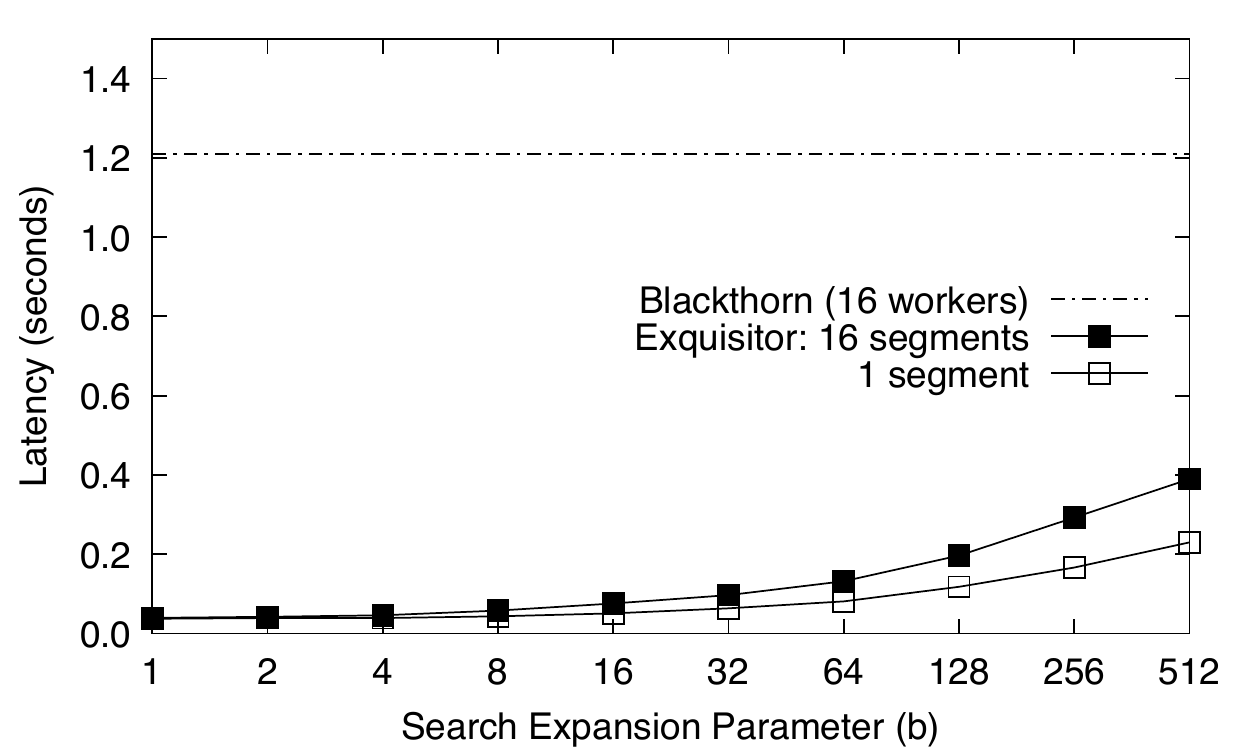}
\caption{Average latency over first 10 rounds of analysis.\hfill \\(Experiment 1: varying $b$; $w = 1$; $S_c=1, 16$).}
\label{fig:e1:t}
\end{figure}

\subsection{Experiment 1: Impact of Search Expansion}
\label{sec:exp:e1}

In this experiment, we explore the impact of the high-dimensional index. The primary parameter in the scoring process is $b$, the number of clusters read and scored. Figure~\ref{fig:e1:p} analyses the impact of $b$ on the precision (fraction of relevant items seen) in each round of the interactive exploration. The $x$-axis shows how many clusters are read for scoring at each round, ranging from $b=1$ to $b=512$ (note the logarithmic scale of the axis), while the $y$-axis shows the average precision across the first 10 rounds of analysis. The figure shows precision for two Exquisitor variants, with $S_c=1$ and $S_c=16$. In both cases, only one worker is used, $w=1$.  For comparison, the figure also shows the average precision for Blackthorn.

%\jz{$\downarrow$ I actually think that the first sentence is unnecessarily negative. My takeaway from the plot is ``considering a moderate number of clusters already matches and exceeds Blackthorn --- awesome --- but it's not like $b=1$ comes out as rubbish, in fact, the one cluster is responsible for 2/3 of the precision, which is remarkable!''}
As Figure~\ref{fig:e1:p} shows, result quality is surprisingly good when scoring only a single cluster in each interaction round, returning about two-thirds of the precision of the state-of-the-art algorithm. As more clusters are considered, quality then improves further. As expected, dividing the $b$ clusters into $S_c=16$ chunks results in better quality, an effect that becomes more and more pronounced as $b$ grows. In particular, with $b=256$,  Exquisitor  returns significantly better results than Blackthorn, even though  Blackthorn  considers every media item in the collection. The reason is that 
%in the Blackthorn system the \textit{mono-modal} media items discussed in Section~\ref{sec:sys} are assigned randomly to workers, where they manage to dominate the \textit{bi-modal} media items; 
by assigning the $b$ relevant clusters to $S_c$ segments,  Exquisitor  is able to emphasize more the bi-modal media items. 
Note that as further clusters are added with Exquisitor ($b=512$ and beyond), the results become more and more similar to the Blackthorn results.

\begin{figure}
	\includegraphics[width=\figwid]{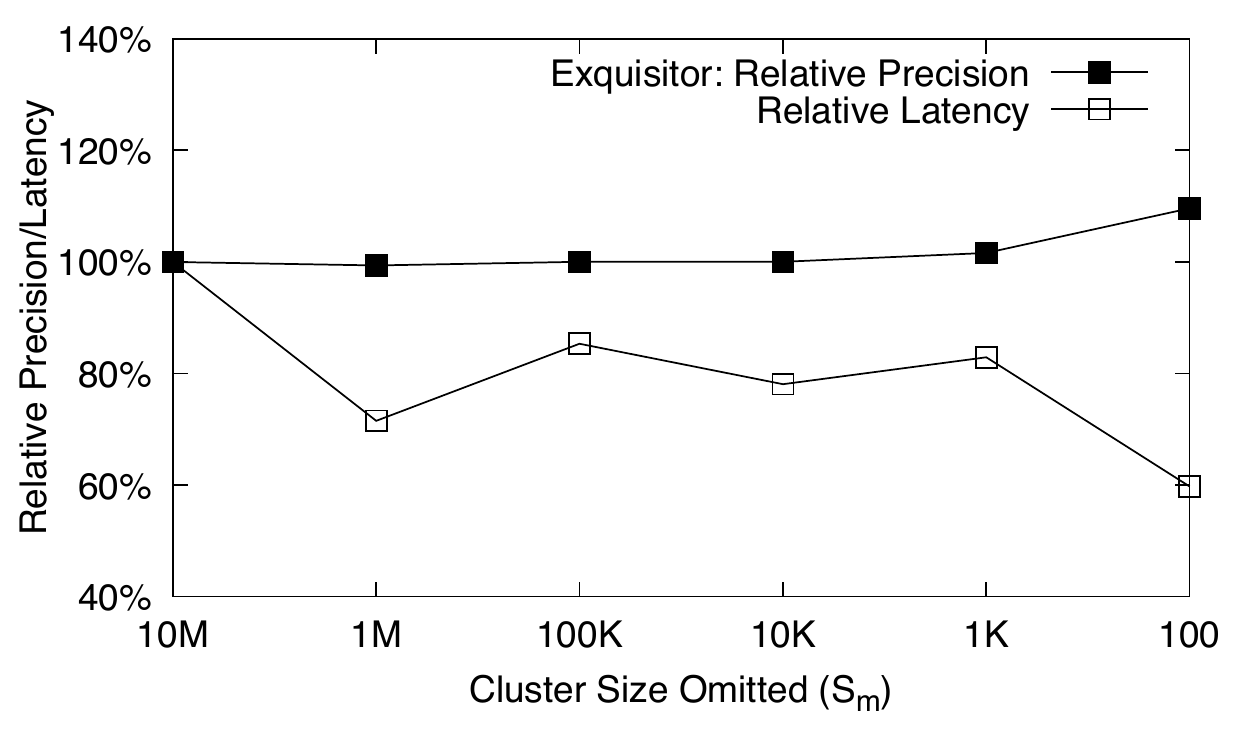}
	\caption{Impact of omitting large clusters on latency and precision. (Experiment 2: varying $S_m$; $b = 256$; $w = 1$; $S_c=16$)}
	\label{fig:e2:tp}
\end{figure}

Figure~\ref{fig:e1:t}, on the other hand, shows the latency per interaction round. 
%As before, the $x$-axis shows how many clusters are read for scoring at each round, while the $y$-axis shows the average latency across the first 10 rounds of analysis.
The figure shows the two Exquisitor variants, with $S_c=1$ and $S_c=16$; in both cases, one worker is used, $w=1$.  For comparison, as before, it  also shows the average latency for Blackthorn (with 16 CPU cores).
%\jz{$\downarrow$ Careful with the ``$w=1$ in Blackthorn = 16x the time'' statement. I haven't tested it exactly, but I know that the lion's share of the 1.2 secs is the merging of the sub-lists, I'd expect a $w=1$ Blackthorn to be on the order of 5-6 secs (I think I tested it briefly in an ancient version of Blackthorn). I think the ``Exquisitor is way more efficient than Blackthorn on both time and CPU resources'' is a strong enough conclusion as it is, you can leave out the ``20 sec Blackthorn'' claim without weakening it IMHO.}
Unsurprisingly, Figure~\ref{fig:e1:t} shows linear growth in latency with respect to $b$ (recall the logarithmic $x$-axis). With $b=256$, each interaction round takes about 0.29 seconds with $S_c=16$, and about 0.17 seconds with $S_c=1$. Both clearly  allow for interactive performance; the remainder of our experiments focus on $b=256$. If  even shorter latency is desired, however, fewer clusters can be read: $b=32$, for example, also gives a good tradeoff between latency and result quality. Recall that this latency is produced using only a single CPU core, meaning that the latency is about 4x better than Blackthorn, with 16x fewer computing cores, for an improvement of about 64x, or nearly two orders of magnitude.

\subsection{Experiment 2: Impact of Data Skew}
\label{sec:exp:e2}

In this experiment, we explore the impact of handling data skew 
%on  latency and precision. 
%As described in Section~\ref{sec:sys:ext}, the first approach to countering data skew is to 
by omitting large clusters from consideration. Figure~\ref{fig:e2:tp} shows the impact of the $S_m$ parameter on both latency and precision. The $x$-axis shows $S_m$, as it is decreased from 10M (no impact, since the largest cluster is 3.5M), to 1M (excluding one cluster in the visual modality and two in the text modality), and then further down to 100 features, where it excludes very large portions of the collection. The $y$-axis shows the relative impact on both precision and latency, compared to the results when all clusters are considered.

Figure~\ref{fig:e2:tp} shows that omitting the three largest clusters ($S_m=$1M) improves latency considerably, without any impact on precision. As more and more clusters are omitted from consideration, however, the impact on either parameter is minimal, until $S_m=100$, where both latency and precision are improved further. While this is an interesting effect that warrants further exploration, and is most likely related again to the tradeoff between bi-modal and mono-modal items, we leave it to future work to analyse it in detail. %In the last experiment we instead use $M=$1M, which omits only the extremely large and uninformative clusters from consideration.

\begin{figure}
\includegraphics[width=\figwid]{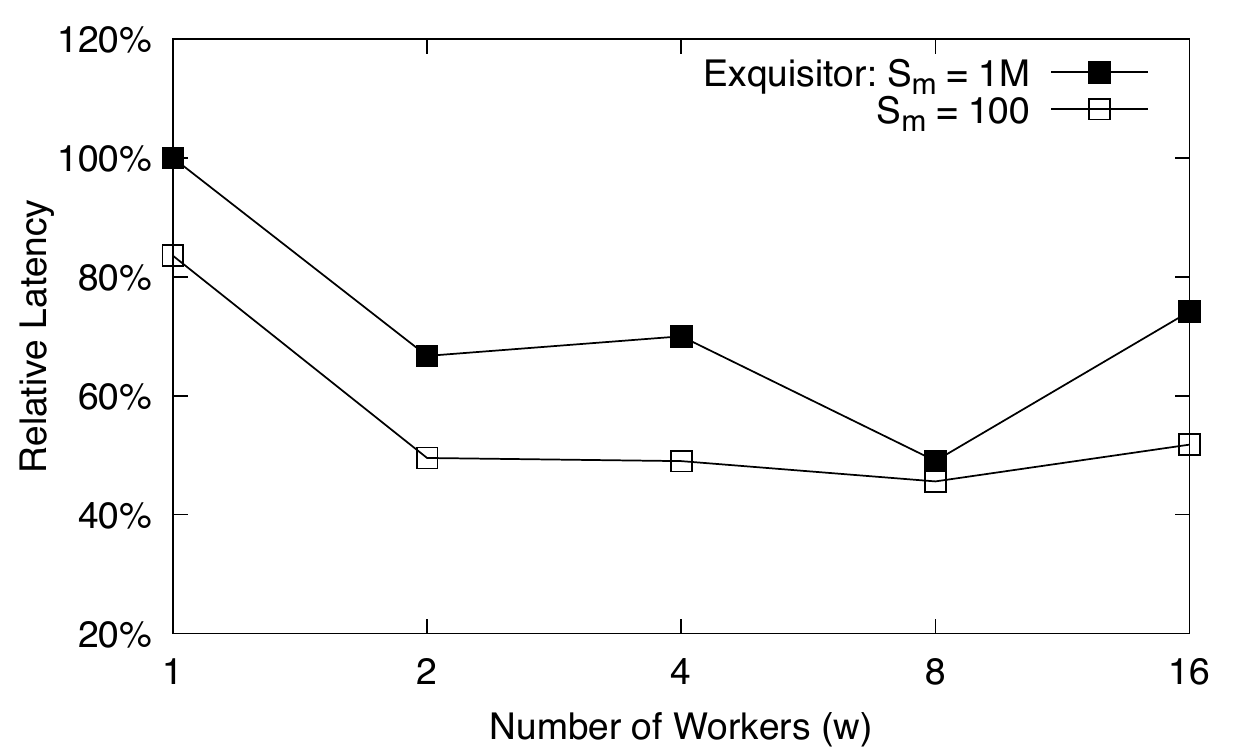}
\caption{Impact of adding workers on latency.\hfill\\ (Experiment 3: varying $w$; $b=256$, $S_m = 1$M, 100; $S_c=16$)}
\label{fig:e3:t}
\end{figure}

%\vfill

\subsection{Experiment 3: Multi-Core Processing}
\label{sec:exp:e3}

The primary parameter in this experiment is $w$, the number of workers applied to the scoring process.  Based on the previous results, we read $b=256$ clusters in $S_c=16$ segments in each iteration, and omit from consideration clusters larger than $S_m=$1M or $S_m=100$, respectively. Note that as workers are added, each worker reads a proportionately smaller share of the $S_c=16$ segments, so precision is not affected by adding workers. We therefore focus on latency.

Figure~\ref{fig:e3:t} shows the evolution of latency, relative to a single worker, as the number of workers (or CPU cores) is varied from $w=1$ to $w=16$. 
As the figure shows, response time is improved somewhat by adding more workers, with latency improved by about 40\% with $w=8$ workers. The reason latency is not improved further is that a)~much of the processing in each interaction round takes place on a single CPU core, including updating the linear SVM model and merging the suggestion lists produced by workers, and b)~the scoring process is already so efficient that it does not benefit further from the added CPU cores.

%\vfill

%\subsection{Summary}
%\label{sec:exp:sum}
%
%To summarize, the results show that Exquisitor improves the state of the art by nearly two orders of magnitude, producing suggestions of better quality in about 0.3 seconds using only a single computation core.  Overall, these results mean that interactive exploration of YFCC100M is now possible on extremely modest hardware.

%!TEX root = ../Exquisitor.tex

\section{Conclusions}
\label{sec:cc}

In this paper, we presented Exquisitor, a new approach for exploratory analysis of very large image collections with modest computational requirements. Exquisitor combines state-of-the-art large-scale interactive learning with a new cluster-based retrieval mechanism, enhancing the relevance capabilities of interactive learning by exploiting the inherent structure of the data. Experiments on YFCC100M, the largest publicly available multimedia collection, show that Exquisitor achieves higher precision and lower latency, with less computational resources, resulting in performance improvements on nearly two orders of magnitude. Additionally, Exquisitor introduces customizability that is, to the best of our knowledge, previously unseen in large-scale interactive learning, by: (i)~allowing a tradeoff between low latency (few clusters) and high quality (many clusters); (ii)~combating data skew by omitting huge (and thus likely nondescript) clusters from consideration; and (iii)~improving latency by adding CPU cores, without any impact on quality.
%\begin{itemize}
%    \item Based on the number of clusters considered each interaction round, Exquisitor can be further customized towards speed (low number of clusters) or precision (high number of clusters).
%    \item The ability to combat data skew by dropping huge, and thus probably non-descript clusters without taking a precision hit.
%    \item Independence of result precision on the way data are split between CPU workers.
%\end{itemize}
In conclusion, Exquisitor provides the best performance on very large collections while being efficient enough to bring large-scale multimedia analytics to standard desktops and laptops, and even high-end mobile devices.

% Bibliography
%\vfill
%\pagebreak
\balance
\bibliographystyle{ACM-Reference-Format}
\bibliography{bibs/n.references}

%%% -*-BibTeX-*-
%%% Do NOT edit. File created by BibTeX with style
%%% ACM-Reference-Format-Journals [18-Jan-2012].

\begin{thebibliography}{78}

%%% ====================================================================
%%% NOTE TO THE USER: you can override these defaults by providing
%%% customized versions of any of these macros before the \bibliography
%%% command.  Each of them MUST provide its own final punctuation,
%%% except for \shownote{}, \showDOI{}, and \showURL{}.  The latter two
%%% do not use final punctuation, in order to avoid confusing it with
%%% the Web address.
%%%
%%% To suppress output of a particular field, define its macro to expand
%%% to an empty string, or better, \unskip, like this:
%%%
%%% \newcommand{\showDOI}[1]{\unskip}   % LaTeX syntax
%%%
%%% \def \showDOI #1{\unskip}           % plain TeX syntax
%%%
%%% ====================================================================

\ifx \showCODEN    \undefined \def \showCODEN     #1{\unskip}     \fi
\ifx \showDOI      \undefined \def \showDOI       #1{#1}\fi
\ifx \showISBNx    \undefined \def \showISBNx     #1{\unskip}     \fi
\ifx \showISBNxiii \undefined \def \showISBNxiii  #1{\unskip}     \fi
\ifx \showISSN     \undefined \def \showISSN      #1{\unskip}     \fi
\ifx \showLCCN     \undefined \def \showLCCN      #1{\unskip}     \fi
\ifx \shownote     \undefined \def \shownote      #1{#1}          \fi
\ifx \showarticletitle \undefined \def \showarticletitle #1{#1}   \fi
\ifx \showURL      \undefined \def \showURL       {\relax}        \fi
% The following commands are used for tagged output and should be
% invisible to TeX
\providecommand\bibfield[2]{#2}
\providecommand\bibinfo[2]{#2}
\providecommand\natexlab[1]{#1}
\providecommand\showeprint[2][]{arXiv:#2}

\bibitem[\protect\citeauthoryear{Abbar, Amer{-}Yahia, Indyk, and
  Mahabadi}{Abbar et~al\mbox{.}}{2013a}]%
        {DBLP:conf/www/AbbarAIM13}
\bibfield{author}{\bibinfo{person}{Sofiane Abbar}, \bibinfo{person}{Sihem
  Amer{-}Yahia}, \bibinfo{person}{Piotr Indyk}, {and} \bibinfo{person}{Sepideh
  Mahabadi}.} \bibinfo{year}{2013}\natexlab{a}.
\newblock \showarticletitle{Real-time recommendation of diverse related
  articles}. In \bibinfo{booktitle}{\emph{Proc. {WWW}}}.
  \bibinfo{publisher}{International World Wide Web Conferences Steering
  Committee/ACM}, \bibinfo{address}{Rio de Janeiro, Brazil},
  \bibinfo{pages}{1--12}.
\newblock


\bibitem[\protect\citeauthoryear{Abbar, Amer{-}Yahia, Indyk, Mahabadi, and
  Varadarajan}{Abbar et~al\mbox{.}}{2013b}]%
        {DBLP:conf/compgeom/AbbarAIMV13}
\bibfield{author}{\bibinfo{person}{Sofiane Abbar}, \bibinfo{person}{Sihem
  Amer{-}Yahia}, \bibinfo{person}{Piotr Indyk}, \bibinfo{person}{Sepideh
  Mahabadi}, {and} \bibinfo{person}{Kasturi~R. Varadarajan}.}
  \bibinfo{year}{2013}\natexlab{b}.
\newblock \showarticletitle{Diverse near neighbor problem}. In
  \bibinfo{booktitle}{\emph{Proc. Symposium on Computational Geometry (SOGG)}}.
  \bibinfo{publisher}{ACM}, \bibinfo{address}{Rio de Janeiro, Brazil},
  \bibinfo{pages}{207--214}.
\newblock


\bibitem[\protect\citeauthoryear{Agarwal, Matou{\v{s}}ek, and Suri}{Agarwal
  et~al\mbox{.}}{1991}]%
        {DBLP:journals/comgeo/AgarwalMS91}
\bibfield{author}{\bibinfo{person}{Pankaj~K. Agarwal},
  \bibinfo{person}{Ji{\v{r}}{\'{\i}} Matou{\v{s}}ek}, {and}
  \bibinfo{person}{Subhash Suri}.} \bibinfo{year}{1991}\natexlab{}.
\newblock \showarticletitle{Farthest Neighbors, Maximum Spanning Trees and
  Related Problems in Higher Dimensions}.
\newblock \bibinfo{journal}{\emph{Comput. Geom.}}  \bibinfo{volume}{1}
  (\bibinfo{year}{1991}), \bibinfo{pages}{189--201}.
\newblock


\bibitem[\protect\citeauthoryear{Amsaleg, J{\'{o}}nsson, and Lejsek}{Amsaleg
  et~al\mbox{.}}{2018}]%
        {DBLP:conf/sisap/AmsalegJL18}
\bibfield{author}{\bibinfo{person}{Laurent Amsaleg},
  \bibinfo{person}{Bj{\"{o}}rn~{\TH}{\'{o}}r J{\'{o}}nsson}, {and}
  \bibinfo{person}{Herwig Lejsek}.} \bibinfo{year}{2018}\natexlab{}.
\newblock \showarticletitle{Scalability of the NV-tree: Three Experiments}. In
  \bibinfo{booktitle}{\emph{Proc. SISAP}}. \bibinfo{publisher}{Springer},
  \bibinfo{address}{Lima, Peru}, \bibinfo{pages}{59--72}.
\newblock


\bibitem[\protect\citeauthoryear{Andoni and Indyk}{Andoni and Indyk}{2006}]%
        {10.1109/FOCS.2006.49}
\bibfield{author}{\bibinfo{person}{Alexandr Andoni} {and}
  \bibinfo{person}{Piotr Indyk}.} \bibinfo{year}{2006}\natexlab{}.
\newblock \showarticletitle{Near-Optimal Hashing Algorithms for Approximate
  Nearest Neighbor in High Dimensions}. In
  \bibinfo{booktitle}{\emph{Proceedings of the IEEE Symposium on the
  Foundations of Computer Science}}. \bibinfo{publisher}{IEEE Computer
  Society}, \bibinfo{address}{Berkeley, CA, USA}, \bibinfo{pages}{459--468}.
\newblock


\bibitem[\protect\citeauthoryear{Babenko and Lempitsky}{Babenko and
  Lempitsky}{2012}]%
        {DBLP:conf/cvpr/BabenkoL12}
\bibfield{author}{\bibinfo{person}{Artem Babenko} {and}
  \bibinfo{person}{Victor~S. Lempitsky}.} \bibinfo{year}{2012}\natexlab{}.
\newblock \showarticletitle{The inverted multi-index}. In
  \bibinfo{booktitle}{\emph{Proceedings of the IEEE International Conference on
  Computer Vision \& Pattern Recognition}}. \bibinfo{publisher}{IEEE},
  \bibinfo{address}{Providence, RI, USA}, \bibinfo{pages}{3069--3076}.
\newblock


\bibitem[\protect\citeauthoryear{Babenko and Lempitsky}{Babenko and
  Lempitsky}{2015}]%
        {DBLP:journals/pami/BabenkoL15}
\bibfield{author}{\bibinfo{person}{Artem Babenko} {and}
  \bibinfo{person}{Victor~S. Lempitsky}.} \bibinfo{year}{2015}\natexlab{}.
\newblock \showarticletitle{The Inverted Multi-Index}.
\newblock \bibinfo{journal}{\emph{IEEE Transactions on Pattern Analysis and
  Machine Intelligence}} \bibinfo{volume}{37}, \bibinfo{number}{6}
  (\bibinfo{year}{2015}), \bibinfo{pages}{1247--1260}.
\newblock


\bibitem[\protect\citeauthoryear{Baranchuk, Babenko, and Malkov}{Baranchuk
  et~al\mbox{.}}{2018}]%
        {DBLP:conf/eccv/BaranchukBM18}
\bibfield{author}{\bibinfo{person}{Dmitry Baranchuk}, \bibinfo{person}{Artem
  Babenko}, {and} \bibinfo{person}{Yury Malkov}.}
  \bibinfo{year}{2018}\natexlab{}.
\newblock \showarticletitle{Revisiting the Inverted Indices for Billion-Scale
  Approximate Nearest Neighbors}. In \bibinfo{booktitle}{\emph{Proc. ECCV}}.
  \bibinfo{publisher}{Springer}, \bibinfo{address}{Munich, Germany},
  \bibinfo{pages}{209--224}.
\newblock


\bibitem[\protect\citeauthoryear{Barroso, Marty, Patterson, and
  Ranganathan}{Barroso et~al\mbox{.}}{2017}]%
        {Barroso:2017:AKM:3069398.3015146}
\bibfield{author}{\bibinfo{person}{Luiz Barroso}, \bibinfo{person}{Mike Marty},
  \bibinfo{person}{David Patterson}, {and} \bibinfo{person}{Parthasarathy
  Ranganathan}.} \bibinfo{year}{2017}\natexlab{}.
\newblock \showarticletitle{Attack of the Killer Microseconds}.
\newblock \bibinfo{journal}{\emph{CACM}} \bibinfo{volume}{60},
  \bibinfo{number}{4} (\bibinfo{date}{March} \bibinfo{year}{2017}),
  \bibinfo{pages}{48--54}.
\newblock


\bibitem[\protect\citeauthoryear{Basri, Hassner, and Zelnik{-}Manor}{Basri
  et~al\mbox{.}}{2011}]%
        {DBLP:journals/pami/BasriHZ11}
\bibfield{author}{\bibinfo{person}{Ronen Basri}, \bibinfo{person}{Tal Hassner},
  {and} \bibinfo{person}{Lihi Zelnik{-}Manor}.}
  \bibinfo{year}{2011}\natexlab{}.
\newblock \showarticletitle{Approximate Nearest Subspace Search}.
\newblock \bibinfo{journal}{\emph{{IEEE} Trans. Pattern Anal. Mach. Intell.}}
  \bibinfo{volume}{33}, \bibinfo{number}{2} (\bibinfo{year}{2011}),
  \bibinfo{pages}{266--278}.
\newblock


\bibitem[\protect\citeauthoryear{Beluch, Genewein, N{\"u}rnberger, and
  K{\"o}hler}{Beluch et~al\mbox{.}}{2018}]%
        {beluch2018power}
\bibfield{author}{\bibinfo{person}{William~H Beluch}, \bibinfo{person}{Tim
  Genewein}, \bibinfo{person}{Andreas N{\"u}rnberger}, {and}
  \bibinfo{person}{Jan~M K{\"o}hler}.} \bibinfo{year}{2018}\natexlab{}.
\newblock \showarticletitle{The power of ensembles for active learning in image
  classification}. In \bibinfo{booktitle}{\emph{Proc. IEEE CVPR}}.
  \bibinfo{publisher}{IEEE Computer Society}, \bibinfo{address}{Salt Lake City,
  UT, USA}, \bibinfo{pages}{9368--9377}.
\newblock


\bibitem[\protect\citeauthoryear{Bj{\"o}rling, Gonzalez, and
  Bonnet}{Bj{\"o}rling et~al\mbox{.}}{2017}]%
        {DBLP:conf/fast/BjorlingGB17}
\bibfield{author}{\bibinfo{person}{Matias Bj{\"o}rling},
  \bibinfo{person}{Javier Gonzalez}, {and} \bibinfo{person}{Philippe Bonnet}.}
  \bibinfo{year}{2017}\natexlab{}.
\newblock \showarticletitle{{LightNVM}: The {Linux} Open-Channel {SSD}
  Subsystem}. In \bibinfo{booktitle}{\emph{Proc. USENIX Conference on File and
  Storage Technologies (FAST)}}. \bibinfo{publisher}{USENIX Association},
  \bibinfo{address}{Santa Clara, CA, USA}, \bibinfo{pages}{359--374}.
\newblock


\bibitem[\protect\citeauthoryear{Chang and Lin}{Chang and Lin}{2011}]%
        {CC01a}
\bibfield{author}{\bibinfo{person}{Chih-Chung Chang} {and}
  \bibinfo{person}{Chih-Jen Lin}.} \bibinfo{year}{2011}\natexlab{}.
\newblock \showarticletitle{{LIBSVM}: A library for support vector machines}.
\newblock \bibinfo{journal}{\emph{ACM Transactions on Intelligent Systems and
  Technology}}  \bibinfo{volume}{2} (\bibinfo{year}{2011}),
  \bibinfo{pages}{27:1--27:27}.
\newblock
Issue 3.
\newblock
\shownote{Software available at
  \url{http://www.csie.ntu.edu.tw/~cjlin/libsvm}.}


\bibitem[\protect\citeauthoryear{Choi, Hauff, Laere, and Thomee}{Choi
  et~al\mbox{.}}{2015}]%
        {451edeadebaa4360865c0527aeafb5c9}
\bibfield{author}{\bibinfo{person}{Jaeyoung Choi}, \bibinfo{person}{Claudia
  Hauff}, \bibinfo{person}{Olivier~Van Laere}, {and} \bibinfo{person}{Bart
  Thomee}.} \bibinfo{year}{2015}\natexlab{}.
\newblock \showarticletitle{The placing task at MediaEval 2015}. In
  \bibinfo{booktitle}{\emph{Proceedings of the MediaEval 2015 Workshop}}.
  \bibinfo{publisher}{CEUR}, \bibinfo{address}{Wurzen, Germany}, 2.
\newblock


\bibitem[\protect\citeauthoryear{Cohn, Ghahramani, and Jordan}{Cohn
  et~al\mbox{.}}{1996}]%
        {Cohn:1996:ALS:1622737.1622744}
\bibfield{author}{\bibinfo{person}{David~A. Cohn}, \bibinfo{person}{Zoubin
  Ghahramani}, {and} \bibinfo{person}{Michael~I. Jordan}.}
  \bibinfo{year}{1996}\natexlab{}.
\newblock \showarticletitle{Active Learning with Statistical Models}.
\newblock \bibinfo{journal}{\emph{JAIR}} \bibinfo{volume}{4},
  \bibinfo{number}{1} (\bibinfo{date}{March} \bibinfo{year}{1996}),
  \bibinfo{pages}{129--145}.
\newblock


\bibitem[\protect\citeauthoryear{Crucianu, Estevez, Oria, and Tarel}{Crucianu
  et~al\mbox{.}}{2008}]%
        {DBLP:journals/imst/CrucianuEOT08}
\bibfield{author}{\bibinfo{person}{Michel Crucianu}, \bibinfo{person}{Daniel
  Estevez}, \bibinfo{person}{Vincent Oria}, {and}
  \bibinfo{person}{Jean{-}Philippe Tarel}.} \bibinfo{year}{2008}\natexlab{}.
\newblock \showarticletitle{Speeding up active relevance feedback with
  approximate \emph{k}NN retrieval for hyperplane queries}.
\newblock \bibinfo{journal}{\emph{Int. J. Imaging Systems and Technology}}
  \bibinfo{volume}{18}, \bibinfo{number}{2-3} (\bibinfo{year}{2008}),
  \bibinfo{pages}{150--159}.
\newblock


\bibitem[\protect\citeauthoryear{Curtin and Gardner}{Curtin and
  Gardner}{2016}]%
        {DBLP:conf/sisap/CurtinG16}
\bibfield{author}{\bibinfo{person}{Ryan~R. Curtin} {and}
  \bibinfo{person}{Andrew~B. Gardner}.} \bibinfo{year}{2016}\natexlab{}.
\newblock \showarticletitle{Fast Approximate Furthest Neighbors with
  Data-Dependent Candidate Selection}. In \bibinfo{booktitle}{\emph{Proc.
  {SISAP}}}. \bibinfo{publisher}{Springer}, \bibinfo{address}{Tokyo, Japan},
  \bibinfo{pages}{221--235}.
\newblock


\bibitem[\protect\citeauthoryear{Datar, Immorlica, Indyk, and Mirrokni}{Datar
  et~al\mbox{.}}{2004}]%
        {Datar:2004:LHS:997817.997857}
\bibfield{author}{\bibinfo{person}{Mayur Datar}, \bibinfo{person}{Nicole
  Immorlica}, \bibinfo{person}{Piotr Indyk}, {and} \bibinfo{person}{Vahab~S.
  Mirrokni}.} \bibinfo{year}{2004}\natexlab{}.
\newblock \showarticletitle{Locality-sensitive hashing scheme based on p-stable
  distributions}. In \bibinfo{booktitle}{\emph{Proc. ACM Symposium on
  Computational Geometry}}. \bibinfo{publisher}{ACM},
  \bibinfo{address}{Brooklyn, NY, USA}, \bibinfo{pages}{253--262}.
\newblock


\bibitem[\protect\citeauthoryear{de~Berg, Cheong, van Kreveld, and
  Overmars}{de~Berg et~al\mbox{.}}{2008}]%
        {DBLP:books/lib/BergCKO08}
\bibfield{author}{\bibinfo{person}{Mark de Berg}, \bibinfo{person}{Otfried
  Cheong}, \bibinfo{person}{Marc~J. van Kreveld}, {and}
  \bibinfo{person}{Mark~H. Overmars}.} \bibinfo{year}{2008}\natexlab{}.
\newblock \bibinfo{booktitle}{\emph{Computational geometry: algorithms and
  applications, 3rd Edition}}.
\newblock \bibinfo{publisher}{Springer}, \bibinfo{address}{Berlin}.
\newblock


\bibitem[\protect\citeauthoryear{{Flickner}, {Sawhney}, {Niblack}, {Ashley}, ,
  {Dom}, {Gorkani}, {Hafner}, {Lee}, {Petkovic}, {Steele}, and
  {Yanker}}{{Flickner} et~al\mbox{.}}{1995}]%
        {410146}
\bibfield{author}{\bibinfo{person}{M. {Flickner}}, \bibinfo{person}{H.
  {Sawhney}}, \bibinfo{person}{W. {Niblack}}, \bibinfo{person}{J. {Ashley}},
  \bibinfo{person}{}, \bibinfo{person}{B. {Dom}}, \bibinfo{person}{M.
  {Gorkani}}, \bibinfo{person}{J. {Hafner}}, \bibinfo{person}{D. {Lee}},
  \bibinfo{person}{D. {Petkovic}}, \bibinfo{person}{D. {Steele}}, {and}
  \bibinfo{person}{P. {Yanker}}.} \bibinfo{year}{1995}\natexlab{}.
\newblock \showarticletitle{Query by image and video content: the QBIC system}.
\newblock \bibinfo{journal}{\emph{Computer}} \bibinfo{volume}{28},
  \bibinfo{number}{9} (\bibinfo{date}{Sep.} \bibinfo{year}{1995}),
  \bibinfo{pages}{23--32}.
\newblock
\showISSN{0018-9162}
\urldef\tempurl%
\url{https://doi.org/10.1109/2.410146}
\showDOI{\tempurl}


\bibitem[\protect\citeauthoryear{Ge, He, Ke, and Sun}{Ge et~al\mbox{.}}{2014}]%
        {DBLP:journals/pami/GeHK014}
\bibfield{author}{\bibinfo{person}{Tiezheng Ge}, \bibinfo{person}{Kaiming He},
  \bibinfo{person}{Qifa Ke}, {and} \bibinfo{person}{Jian Sun}.}
  \bibinfo{year}{2014}\natexlab{}.
\newblock \showarticletitle{Optimized Product Quantization}.
\newblock \bibinfo{journal}{\emph{IEEE Transactions on Pattern Analysis and
  Machine Intelligence}} \bibinfo{volume}{36}, \bibinfo{number}{4}
  (\bibinfo{year}{2014}), \bibinfo{pages}{744--755}.
\newblock


\bibitem[\protect\citeauthoryear{Gionis, Indyk, and Motwani}{Gionis
  et~al\mbox{.}}{1999}]%
        {Gionis:1999:SSH:645925.671516}
\bibfield{author}{\bibinfo{person}{Aristides Gionis}, \bibinfo{person}{Piotr
  Indyk}, {and} \bibinfo{person}{Rajeev Motwani}.}
  \bibinfo{year}{1999}\natexlab{}.
\newblock \showarticletitle{Similarity Search in High Dimensions via Hashing}.
  In \bibinfo{booktitle}{\emph{Proceedings of the International Conference on
  Very Large Data Bases}}. \bibinfo{publisher}{Morgan Kaufmann},
  \bibinfo{address}{Edinburgh, Scotland}, \bibinfo{pages}{518--529}.
\newblock


\bibitem[\protect\citeauthoryear{Gudmundsson, Amsaleg, and
  J\'{o}nsson}{Gudmundsson et~al\mbox{.}}{2012}]%
        {Gudmundsson:2012:IST:2261008.2261018}
\bibfield{author}{\bibinfo{person}{Gylfi~\TH\'{o}r Gudmundsson},
  \bibinfo{person}{Laurent Amsaleg}, {and} \bibinfo{person}{Bj\"{o}rn~\TH\'{o}r
  J\'{o}nsson}.} \bibinfo{year}{2012}\natexlab{}.
\newblock \showarticletitle{Impact of Storage Technology on the Efficiency of
  Cluster-based High-dimensional Index Creation}. In
  \bibinfo{booktitle}{\emph{Proc. International Conference on Database Systems
  for Advanced Applications (DASFAA)}}. \bibinfo{publisher}{Springer},
  \bibinfo{address}{Busan, South Korea}, \bibinfo{pages}{53--64}.
\newblock


\bibitem[\protect\citeauthoryear{Gudmundsson, Amsaleg, J\'{o}nsson, and
  Franklin}{Gudmundsson et~al\mbox{.}}{2017}]%
        {GudmundssonMMSys17}
\bibfield{author}{\bibinfo{person}{Gylfi~\TH\'{o}r Gudmundsson},
  \bibinfo{person}{Laurent Amsaleg}, \bibinfo{person}{Bj\"{o}rn~\TH\'{o}r
  J\'{o}nsson}, {and} \bibinfo{person}{Michael~J. Franklin}.}
  \bibinfo{year}{2017}\natexlab{}.
\newblock \showarticletitle{Towards Engineering a Web-Scale Multimedia Service:
  A Case Study Using {Spark}}. In \bibinfo{booktitle}{\emph{Proc. ACM
  Multimedia Systems Conference (MMSys)}}. \bibinfo{publisher}{ACM},
  \bibinfo{address}{Taipei, Taiwan}, \bibinfo{pages}{1--12}.
\newblock


\bibitem[\protect\citeauthoryear{Gudmundsson, J\'{o}nsson, and
  Amsaleg}{Gudmundsson et~al\mbox{.}}{2010}]%
        {Gudmundsson:2010:LPS:1878137.1878145}
\bibfield{author}{\bibinfo{person}{Gylfi~\TH\'{o}r Gudmundsson},
  \bibinfo{person}{Bj\"{o}rn~\TH\'{o}r J\'{o}nsson}, {and}
  \bibinfo{person}{Laurent Amsaleg}.} \bibinfo{year}{2010}\natexlab{}.
\newblock \showarticletitle{A Large-scale Performance Study of Cluster-based
  High-dimensional Indexing}. In \bibinfo{booktitle}{\emph{Proc. International
  Workshop on Very-large-scale Multimedia Corpus, Mining and Retrieval
  (VLS-MCMR)}}. \bibinfo{publisher}{ACM}, \bibinfo{address}{Firenze, Italy},
  \bibinfo{pages}{31--36}.
\newblock


\bibitem[\protect\citeauthoryear{Heo, Lin, and Yoon}{Heo et~al\mbox{.}}{2014}]%
        {DBLP:conf/cvpr/HeoLY14}
\bibfield{author}{\bibinfo{person}{Jae{-}Pil Heo}, \bibinfo{person}{Zhe Lin},
  {and} \bibinfo{person}{Sung{-}Eui Yoon}.} \bibinfo{year}{2014}\natexlab{}.
\newblock \showarticletitle{Distance Encoded Product Quantization}. In
  \bibinfo{booktitle}{\emph{Proceedings of the IEEE International Conference on
  Computer Vision \& Pattern Recognition}}. \bibinfo{publisher}{IEEE Computer
  Society}, \bibinfo{address}{Columbus, OH, USA}, \bibinfo{pages}{2139--2146}.
\newblock


\bibitem[\protect\citeauthoryear{Huang, Dagli, Rajaram, Chang, Mandel, Poliner,
  and Ellis}{Huang et~al\mbox{.}}{2008}]%
        {huang08}
\bibfield{author}{\bibinfo{person}{T.S. Huang}, \bibinfo{person}{C.K. Dagli},
  \bibinfo{person}{S. Rajaram}, \bibinfo{person}{E.Y. Chang},
  \bibinfo{person}{M.I. Mandel}, \bibinfo{person}{Graham~E. Poliner}, {and}
  \bibinfo{person}{D.P.W. Ellis}.} \bibinfo{year}{2008}\natexlab{}.
\newblock \showarticletitle{Active Learning for Interactive Multimedia
  Retrieval}.
\newblock \bibinfo{journal}{\emph{Proc. IEEE}} \bibinfo{volume}{96},
  \bibinfo{number}{4} (\bibinfo{year}{2008}), \bibinfo{pages}{648--667}.
\newblock
\showISSN{0018-9219}
\urldef\tempurl%
\url{https://doi.org/10.1109/JPROC.2008.916364}
\showDOI{\tempurl}


\bibitem[\protect\citeauthoryear{{Huijser} and v.~{Gemert}}{{Huijser} and
  v.~{Gemert}}{2017}]%
        {8237827}
\bibfield{author}{\bibinfo{person}{M. {Huijser}} {and} \bibinfo{person}{J.~C.
  v. {Gemert}}.} \bibinfo{year}{2017}\natexlab{}.
\newblock \showarticletitle{Active Decision Boundary Annotation with Deep
  Generative Models}. In \bibinfo{booktitle}{\emph{Proc. IEEE International
  Conference on Computer Vision (ICCV)}}. \bibinfo{publisher}{IEEE Computer
  Society}, \bibinfo{address}{Venice, Italy}, \bibinfo{pages}{5296--5305}.
\newblock


\bibitem[\protect\citeauthoryear{Indyk}{Indyk}{2003}]%
        {DBLP:conf/soda/Indyk03}
\bibfield{author}{\bibinfo{person}{Piotr Indyk}.}
  \bibinfo{year}{2003}\natexlab{}.
\newblock \showarticletitle{Better algorithms for high-dimensional proximity
  problems via asymmetric embeddings}. In \bibinfo{booktitle}{\emph{Proc.
  ACM-SIAM Symposium on Discrete Algorithms (SODA)}}.
  \bibinfo{publisher}{ACM/SIAM}, \bibinfo{address}{Baltimore, MD, USA},
  \bibinfo{pages}{539--545}.
\newblock


\bibitem[\protect\citeauthoryear{{Iscen}, {Furon}, {Gripon}, {Rabbat}, and
  {Jégou}}{{Iscen} et~al\mbox{.}}{2018}]%
        {7870636}
\bibfield{author}{\bibinfo{person}{A. {Iscen}}, \bibinfo{person}{T. {Furon}},
  \bibinfo{person}{V. {Gripon}}, \bibinfo{person}{M. {Rabbat}}, {and}
  \bibinfo{person}{H. {Jégou}}.} \bibinfo{year}{2018}\natexlab{}.
\newblock \showarticletitle{Memory Vectors for Similarity Search in
  High-Dimensional Spaces}.
\newblock \bibinfo{journal}{\emph{IEEE Transactions on Big Data}}
  \bibinfo{volume}{4}, \bibinfo{number}{1} (\bibinfo{date}{March}
  \bibinfo{year}{2018}), \bibinfo{pages}{65--77}.
\newblock
\showISSN{2332-7790}
\urldef\tempurl%
\url{https://doi.org/10.1109/TBDATA.2017.2677964}
\showDOI{\tempurl}


\bibitem[\protect\citeauthoryear{Jain, Kulis, and Grauman}{Jain
  et~al\mbox{.}}{2008}]%
        {DBLP:conf/cvpr/JainKG08}
\bibfield{author}{\bibinfo{person}{Prateek Jain}, \bibinfo{person}{Brian
  Kulis}, {and} \bibinfo{person}{Kristen Grauman}.}
  \bibinfo{year}{2008}\natexlab{}.
\newblock \showarticletitle{Fast image search for learned metrics}. In
  \bibinfo{booktitle}{\emph{Proceedings of the IEEE International Conference on
  Computer Vision \& Pattern Recognition}}. \bibinfo{publisher}{IEEE Computer
  Society}, \bibinfo{address}{Anchorage, AK, USA}.
\newblock


\bibitem[\protect\citeauthoryear{Jain, Vijayanarasimhan, and Grauman}{Jain
  et~al\mbox{.}}{2010}]%
        {DBLP:conf/nips/JainVG10}
\bibfield{author}{\bibinfo{person}{Prateek Jain}, \bibinfo{person}{Sudheendra
  Vijayanarasimhan}, {and} \bibinfo{person}{Kristen Grauman}.}
  \bibinfo{year}{2010}\natexlab{}.
\newblock \showarticletitle{Hashing Hyperplane Queries to Near Points with
  Applications to Large-Scale Active Learning}. In
  \bibinfo{booktitle}{\emph{Proc. Conference on Neural Information Processing
  Systems (NIPS)}}. \bibinfo{publisher}{Curran Associates, Inc.},
  \bibinfo{address}{Vancouver, BC, Canada}, \bibinfo{pages}{928--936}.
\newblock


\bibitem[\protect\citeauthoryear{J{\'e}gou, Douze, and Schmid}{J{\'e}gou
  et~al\mbox{.}}{2011}]%
        {jegou:inria-00514462}
\bibfield{author}{\bibinfo{person}{Herv{\'e} J{\'e}gou},
  \bibinfo{person}{Matthijs Douze}, {and} \bibinfo{person}{Cordelia Schmid}.}
  \bibinfo{year}{2011}\natexlab{}.
\newblock \showarticletitle{Product Quantization for Nearest Neighbor Search}.
\newblock \bibinfo{journal}{\emph{IEEE Transactions on Pattern Analysis and
  Machine Intelligence}} \bibinfo{volume}{33}, \bibinfo{number}{1}
  (\bibinfo{year}{2011}), \bibinfo{pages}{117--128}.
\newblock


\bibitem[\protect\citeauthoryear{Jiang, Yu, Meng, Yang, Mitamura, and
  Hauptmann}{Jiang et~al\mbox{.}}{2015}]%
        {Jiang:2015:FAC:2733373.2806237}
\bibfield{author}{\bibinfo{person}{Lu Jiang}, \bibinfo{person}{Shoou-I Yu},
  \bibinfo{person}{Deyu Meng}, \bibinfo{person}{Yi Yang},
  \bibinfo{person}{Teruko Mitamura}, {and} \bibinfo{person}{Alexander~G.
  Hauptmann}.} \bibinfo{year}{2015}\natexlab{}.
\newblock \showarticletitle{Fast and Accurate Content-based Semantic Search in
  100M Internet Videos}. In \bibinfo{booktitle}{\emph{Proc. ACM Multimedia}}.
  \bibinfo{publisher}{ACM}, \bibinfo{address}{Brisbane, Australia},
  \bibinfo{pages}{49--58}.
\newblock


\bibitem[\protect\citeauthoryear{Jin, Hu, Lin, Zhang, Lin, Cai, and Li}{Jin
  et~al\mbox{.}}{2013}]%
        {DBLP:conf/iccv/JinHLZLCL13}
\bibfield{author}{\bibinfo{person}{Zhongming Jin}, \bibinfo{person}{Yao Hu},
  \bibinfo{person}{Yue Lin}, \bibinfo{person}{Debing Zhang},
  \bibinfo{person}{Shiding Lin}, \bibinfo{person}{Deng Cai}, {and}
  \bibinfo{person}{Xuelong Li}.} \bibinfo{year}{2013}\natexlab{}.
\newblock \showarticletitle{Complementary Projection Hashing}. In
  \bibinfo{booktitle}{\emph{Proceedings of the IEEE International Conference on
  Computer Vision}}. \bibinfo{publisher}{IEEE Computer Society},
  \bibinfo{address}{Barcelona, Spain}, \bibinfo{pages}{257--264}.
\newblock


\bibitem[\protect\citeauthoryear{Joly and Buisson}{Joly and Buisson}{2008}]%
        {Joly:2008:PML:1459359.1459388}
\bibfield{author}{\bibinfo{person}{Alexis Joly} {and} \bibinfo{person}{Olivier
  Buisson}.} \bibinfo{year}{2008}\natexlab{}.
\newblock \showarticletitle{A posteriori multi-probe locality sensitive
  hashing}. In \bibinfo{booktitle}{\emph{Proceedings of the ACM International
  Conference on Multimedia}}. \bibinfo{publisher}{ACM},
  \bibinfo{address}{Vancouver, BC, Canada}, \bibinfo{pages}{209--218}.
\newblock


\bibitem[\protect\citeauthoryear{Kalantidis and Avrithis}{Kalantidis and
  Avrithis}{2014}]%
        {DBLP:conf/cvpr/KalantidisA14}
\bibfield{author}{\bibinfo{person}{Yannis~S. Kalantidis} {and}
  \bibinfo{person}{Yannis Avrithis}.} \bibinfo{year}{2014}\natexlab{}.
\newblock \showarticletitle{Locally Optimized Product Quantization for
  Approximate Nearest Neighbor Search}. In
  \bibinfo{booktitle}{\emph{Proceedings of the IEEE International Conference on
  Computer Vision \& Pattern Recognition}}. \bibinfo{publisher}{IEEE Computer
  Society}, \bibinfo{address}{Columbus, OH, USA}, \bibinfo{pages}{2329--2336}.
\newblock


\bibitem[\protect\citeauthoryear{Kelly}{Kelly}{2019}]%
        {forbes}
\bibfield{author}{\bibinfo{person}{Gordon Kelly}.} \bibinfo{year}{Jan 15,
  2019}\natexlab{}.
\newblock \bibinfo{title}{Samsung Holding Back Galaxy S10's RAM And Storage?}
\newblock   (\bibinfo{year}{Jan 15, 2019}).
\newblock
\urldef\tempurl%
\url{https://www.forbes.com/sites/gordonkelly/2019/01/15/samsung-galaxy-s10-upgrade-release-date-cost-specs-ram-storage-galaxy-s9-note9/}
\showURL{%
\tempurl}


\bibitem[\protect\citeauthoryear{Kovashka, Parikh, and Grauman}{Kovashka
  et~al\mbox{.}}{2015}]%
        {Kovashka2015}
\bibfield{author}{\bibinfo{person}{Adriana Kovashka}, \bibinfo{person}{Devi
  Parikh}, {and} \bibinfo{person}{Kristen Grauman}.}
  \bibinfo{year}{2015}\natexlab{}.
\newblock \showarticletitle{WhittleSearch: Interactive Image Search with
  Relative Attribute Feedback}.
\newblock \bibinfo{journal}{\emph{International Journal of Computer Vision}}
  \bibinfo{volume}{115}, \bibinfo{number}{2} (\bibinfo{date}{01 Nov}
  \bibinfo{year}{2015}), \bibinfo{pages}{185--210}.
\newblock
\showISSN{1573-1405}
\urldef\tempurl%
\url{https://doi.org/10.1007/s11263-015-0814-0}
\showDOI{\tempurl}


\bibitem[\protect\citeauthoryear{Larson, Soleymani, Serdyukov, Rudinac,
  Wartena, Murdock, Friedland, Ordelman, and Jones}{Larson
  et~al\mbox{.}}{2011}]%
        {Larson:2011:ATG:1991996.1992047}
\bibfield{author}{\bibinfo{person}{Martha Larson}, \bibinfo{person}{Mohammad
  Soleymani}, \bibinfo{person}{Pavel Serdyukov}, \bibinfo{person}{Stevan
  Rudinac}, \bibinfo{person}{Christian Wartena}, \bibinfo{person}{Vanessa
  Murdock}, \bibinfo{person}{Gerald Friedland}, \bibinfo{person}{Roeland
  Ordelman}, {and} \bibinfo{person}{Gareth J.~F. Jones}.}
  \bibinfo{year}{2011}\natexlab{}.
\newblock \showarticletitle{Automatic Tagging and Geotagging in Video
  Collections and Communities}. In \bibinfo{booktitle}{\emph{Proceedings of the
  1st ACM International Conference on Multimedia Retrieval}}
  \emph{(\bibinfo{series}{ICMR '11})}. \bibinfo{publisher}{ACM},
  \bibinfo{address}{New York, NY, USA}, Article \bibinfo{articleno}{51},
  \bibinfo{numpages}{8}~pages.
\newblock
\showISBNx{978-1-4503-0336-1}
\urldef\tempurl%
\url{https://doi.org/10.1145/1991996.1992047}
\showDOI{\tempurl}


\bibitem[\protect\citeauthoryear{Lejsek, {\'{A}}smundsson, J{\'{o}}nsson, and
  Amsaleg}{Lejsek et~al\mbox{.}}{2006}]%
        {DBLP:conf/mm/LejsekAJA06a}
\bibfield{author}{\bibinfo{person}{Herwig Lejsek},
  \bibinfo{person}{Fri{\dh}rik~Hei{\dh}ar {\'{A}}smundsson},
  \bibinfo{person}{Bj{\"{o}}rn~{\TH}{\'{o}}r J{\'{o}}nsson}, {and}
  \bibinfo{person}{Laurent Amsaleg}.} \bibinfo{year}{2006}\natexlab{}.
\newblock \showarticletitle{Scalability of local image descriptors: a
  comparative study}. In \bibinfo{booktitle}{\emph{Proc. ACM Multimedia}}.
  \bibinfo{publisher}{ACM}, \bibinfo{address}{Santa Barbara, CA, USA},
  \bibinfo{pages}{589--598}.
\newblock


\bibitem[\protect\citeauthoryear{Lejsek, \'Asmun{\dh}sson, J\'onsson, and
  Amsaleg}{Lejsek et~al\mbox{.}}{2009}]%
        {10.1109/TPAMI.2008.130}
\bibfield{author}{\bibinfo{person}{Herwig Lejsek},
  \bibinfo{person}{Fridrik~Heidar \'Asmun{\dh}sson},
  \bibinfo{person}{Bjorn~{\TH}or J\'onsson}, {and} \bibinfo{person}{Laurent
  Amsaleg}.} \bibinfo{year}{2009}\natexlab{}.
\newblock \showarticletitle{{NV-Tree}: An Efficient Disk-Based Index for
  Approximate Search in Very Large High-Dimensional Collections}.
\newblock \bibinfo{journal}{\emph{IEEE Transactions on Pattern Analysis and
  Machine Intelligence}} \bibinfo{volume}{31}, \bibinfo{number}{5}
  (\bibinfo{year}{2009}), \bibinfo{pages}{869--883}.
\newblock


\bibitem[\protect\citeauthoryear{Lejsek, J\'{o}nsson, and Amsaleg}{Lejsek
  et~al\mbox{.}}{2011}]%
        {Lejsek:2011:NNN:1991996.1992050}
\bibfield{author}{\bibinfo{person}{Herwig Lejsek},
  \bibinfo{person}{Bj\"{o}rn~{\TH}\'or J\'{o}nsson}, {and}
  \bibinfo{person}{Laurent Amsaleg}.} \bibinfo{year}{2011}\natexlab{}.
\newblock \showarticletitle{{NV-Tree}: nearest neighbors at the billion scale}.
  In \bibinfo{booktitle}{\emph{Proceedings of the ACM International Conference
  on Multimedia Retrieval}}. \bibinfo{publisher}{ACM},
  \bibinfo{address}{Trento, Italy}, Article \bibinfo{articleno}{54},
  \bibinfo{numpages}{8}~pages.
\newblock


\bibitem[\protect\citeauthoryear{{Lokoč}, {Bailer}, {Schoeffmann}, {Muenzer},
  and {Awad}}{{Lokoč} et~al\mbox{.}}{2018}]%
        {8352047}
\bibfield{author}{\bibinfo{person}{J. {Lokoč}}, \bibinfo{person}{W. {Bailer}},
  \bibinfo{person}{K. {Schoeffmann}}, \bibinfo{person}{B. {Muenzer}}, {and}
  \bibinfo{person}{G. {Awad}}.} \bibinfo{year}{2018}\natexlab{}.
\newblock \showarticletitle{On Influential Trends in Interactive Video
  Retrieval: Video Browser Showdown 2015–2017}.
\newblock \bibinfo{journal}{\emph{IEEE Transactions on Multimedia}}
  \bibinfo{volume}{20}, \bibinfo{number}{12} (\bibinfo{date}{Dec}
  \bibinfo{year}{2018}), \bibinfo{pages}{3361--3376}.
\newblock
\showISSN{1520-9210}
\urldef\tempurl%
\url{https://doi.org/10.1109/TMM.2018.2830110}
\showDOI{\tempurl}


\bibitem[\protect\citeauthoryear{Lv, Josephson, Wang, Charikar, and Li}{Lv
  et~al\mbox{.}}{2007}]%
        {Lv:2007:MLE:1325851.1325958}
\bibfield{author}{\bibinfo{person}{Qin Lv}, \bibinfo{person}{William
  Josephson}, \bibinfo{person}{Zhe Wang}, \bibinfo{person}{Moses Charikar},
  {and} \bibinfo{person}{Kai Li}.} \bibinfo{year}{2007}\natexlab{}.
\newblock \showarticletitle{Multi-probe {LSH}: efficient indexing for
  high-dimensional similarity search}. In \bibinfo{booktitle}{\emph{Proceedings
  of the International Conference on Very Large Data Bases}}.
  \bibinfo{publisher}{VLDB Endowment}, \bibinfo{address}{Vienna, Austria},
  \bibinfo{pages}{950--961}.
\newblock


\bibitem[\protect\citeauthoryear{Mironică, Ionescu, Uijlings, and
  Sebe}{Mironică et~al\mbox{.}}{2016}]%
        {MIRONICA201638}
\bibfield{author}{\bibinfo{person}{Ionuţ Mironică}, \bibinfo{person}{Bogdan
  Ionescu}, \bibinfo{person}{Jasper Uijlings}, {and} \bibinfo{person}{Nicu
  Sebe}.} \bibinfo{year}{2016}\natexlab{}.
\newblock \showarticletitle{Fisher Kernel Temporal Variation-based Relevance
  Feedback for video retrieval}.
\newblock \bibinfo{journal}{\emph{Computer Vision and Image Understanding}}
  \bibinfo{volume}{143} (\bibinfo{year}{2016}), \bibinfo{pages}{38 -- 51}.
\newblock
\showISSN{1077-3142}
\urldef\tempurl%
\url{https://doi.org/10.1016/j.cviu.2015.10.005}
\showDOI{\tempurl}
\newblock
\shownote{Inference and Learning of Graphical Models\: Theory and Applications
  in Computer Vision and Image Analysis.}


\bibitem[\protect\citeauthoryear{{North}}{{North}}{2006}]%
        {1626178}
\bibfield{author}{\bibinfo{person}{C. {North}}.}
  \bibinfo{year}{2006}\natexlab{}.
\newblock \showarticletitle{Toward measuring visualization insight}.
\newblock \bibinfo{journal}{\emph{IEEE Computer Graphics and Applications}}
  \bibinfo{volume}{26}, \bibinfo{number}{3} (\bibinfo{date}{May}
  \bibinfo{year}{2006}), \bibinfo{pages}{6--9}.
\newblock
\showISSN{0272-1716}
\urldef\tempurl%
\url{https://doi.org/10.1109/MCG.2006.70}
\showDOI{\tempurl}


\bibitem[\protect\citeauthoryear{Pagh, Silvestri, Sivertsen, and Skala}{Pagh
  et~al\mbox{.}}{2015}]%
        {DBLP:conf/sisap/PaghSSS15}
\bibfield{author}{\bibinfo{person}{Rasmus Pagh}, \bibinfo{person}{Francesco
  Silvestri}, \bibinfo{person}{Johan Sivertsen}, {and} \bibinfo{person}{Matthew
  Skala}.} \bibinfo{year}{2015}\natexlab{}.
\newblock \showarticletitle{Approximate Furthest Neighbor in High Dimensions}.
  In \bibinfo{booktitle}{\emph{Proc. {SISAP}}}, Vol.~\bibinfo{volume}{9371}.
  \bibinfo{publisher}{Springer}, \bibinfo{address}{Glasgow, Scotland},
  \bibinfo{pages}{3--14}.
\newblock


\bibitem[\protect\citeauthoryear{Pagh, Silvestri, Sivertsen, and Skala}{Pagh
  et~al\mbox{.}}{2017}]%
        {DBLP:journals/is/PaghSSS17}
\bibfield{author}{\bibinfo{person}{Rasmus Pagh}, \bibinfo{person}{Francesco
  Silvestri}, \bibinfo{person}{Johan Sivertsen}, {and} \bibinfo{person}{Matthew
  Skala}.} \bibinfo{year}{2017}\natexlab{}.
\newblock \showarticletitle{Approximate furthest neighbor with application to
  annulus query}.
\newblock \bibinfo{journal}{\emph{Inf. Syst.}}  \bibinfo{volume}{64}
  (\bibinfo{year}{2017}), \bibinfo{pages}{152--162}.
\newblock


\bibitem[\protect\citeauthoryear{Paulev{\'{e}}, J{\'{e}}gou, and
  Amsaleg}{Paulev{\'{e}} et~al\mbox{.}}{2010}]%
        {DBLP:journals/prl/PauleveJA10}
\bibfield{author}{\bibinfo{person}{Lo{\"{\i}}c Paulev{\'{e}}},
  \bibinfo{person}{Herv{\'{e}} J{\'{e}}gou}, {and} \bibinfo{person}{Laurent
  Amsaleg}.} \bibinfo{year}{2010}\natexlab{}.
\newblock \showarticletitle{Locality sensitive hashing: {A} comparison of hash
  function types and querying mechanisms}.
\newblock \bibinfo{journal}{\emph{Pattern Recognition Letters}}
  \bibinfo{volume}{31}, \bibinfo{number}{11} (\bibinfo{year}{2010}),
  \bibinfo{pages}{1348--1358}.
\newblock


\bibitem[\protect\citeauthoryear{Philbin, Chum, Isard, Sivic, and
  Zisserman}{Philbin et~al\mbox{.}}{2008}]%
        {4587635}
\bibfield{author}{\bibinfo{person}{James Philbin}, \bibinfo{person}{Ondra
  Chum}, \bibinfo{person}{Michael Isard}, \bibinfo{person}{Josef Sivic}, {and}
  \bibinfo{person}{Andrew Zisserman}.} \bibinfo{year}{2008}\natexlab{}.
\newblock \showarticletitle{Lost in quantization: Improving particular object
  retrieval in large scale image databases}. In
  \bibinfo{booktitle}{\emph{Proceedings of the IEEE International Conference on
  Computer Vision \& Pattern Recognition}}. \bibinfo{publisher}{IEEE Computer
  Society}, \bibinfo{address}{Anchorage, AK, USA}.
\newblock


\bibitem[\protect\citeauthoryear{{\v R}eh{\r u}{\v r}ek and Sojka}{{\v R}eh{\r
  u}{\v r}ek and Sojka}{2010}]%
        {rehurek_lrec}
\bibfield{author}{\bibinfo{person}{Radim {\v R}eh{\r u}{\v r}ek} {and}
  \bibinfo{person}{Petr Sojka}.} \bibinfo{year}{2010}\natexlab{}.
\newblock \showarticletitle{{Software Framework for Topic Modelling with Large
  Corpora}}. In \bibinfo{booktitle}{\emph{{Proceedings of the LREC 2010
  Workshop on New Challenges for NLP Frameworks}}}. \bibinfo{publisher}{ELRA},
  \bibinfo{address}{Valletta, Malta}, \bibinfo{pages}{45--50}.
\newblock
\newblock
\shownote{\url{http://is.muni.cz/publication/884893/en}.}


\bibitem[\protect\citeauthoryear{Rudinac, Chua, Diaz-Ferreyra, Friedland,
  Gornostaja, Huet, Kaptein, Lind{\'e}n, Moens, Peltonen, Redi, Schedl, Shamma,
  Smeaton, and Xie}{Rudinac et~al\mbox{.}}{2018}]%
        {10.1007/978-3-319-73603-7_51}
\bibfield{author}{\bibinfo{person}{Stevan Rudinac}, \bibinfo{person}{Tat-Seng
  Chua}, \bibinfo{person}{Nicolas Diaz-Ferreyra}, \bibinfo{person}{Gerald
  Friedland}, \bibinfo{person}{Tatjana Gornostaja}, \bibinfo{person}{Benoit
  Huet}, \bibinfo{person}{Rianne Kaptein}, \bibinfo{person}{Krister
  Lind{\'e}n}, \bibinfo{person}{Marie-Francine Moens}, \bibinfo{person}{Jaakko
  Peltonen}, \bibinfo{person}{Miriam Redi}, \bibinfo{person}{Markus Schedl},
  \bibinfo{person}{David~A. Shamma}, \bibinfo{person}{Alan Smeaton}, {and}
  \bibinfo{person}{Lexing Xie}.} \bibinfo{year}{2018}\natexlab{}.
\newblock \showarticletitle{Rethinking Summarization and Storytelling for
  Modern Social Multimedia}. In \bibinfo{booktitle}{\emph{MultiMedia
  Modeling}}, \bibfield{editor}{\bibinfo{person}{Klaus Schoeffmann},
  \bibinfo{person}{Thanarat~H. Chalidabhongse}, \bibinfo{person}{Chong~Wah
  Ngo}, \bibinfo{person}{Supavadee Aramvith}, \bibinfo{person}{Noel~E.
  O'Connor}, \bibinfo{person}{Yo-Sung Ho}, \bibinfo{person}{Moncef Gabbouj},
  {and} \bibinfo{person}{Ahmed Elgammal}} (Eds.). \bibinfo{publisher}{Springer
  International Publishing}, \bibinfo{address}{Cham},
  \bibinfo{pages}{632--644}.
\newblock
\showISBNx{978-3-319-73603-7}


\bibitem[\protect\citeauthoryear{{Rui}, {Huang}, and {Mehrotra}}{{Rui}
  et~al\mbox{.}}{1997}]%
        {638621}
\bibfield{author}{\bibinfo{person}{Y. {Rui}}, \bibinfo{person}{T.~S. {Huang}},
  {and} \bibinfo{person}{S. {Mehrotra}}.} \bibinfo{year}{1997}\natexlab{}.
\newblock \showarticletitle{Content-based image retrieval with relevance
  feedback in {MARS}}. In \bibinfo{booktitle}{\emph{Proc. International
  Conference on Image Processing (ICIP)}}. \bibinfo{publisher}{IEEE Computer
  Society}, \bibinfo{address}{Santa Barbara, CA, USA},
  \bibinfo{pages}{815--818}.
\newblock


\bibitem[\protect\citeauthoryear{Russakovsky, Deng, Su, Krause, Satheesh, Ma,
  Huang, Karpathy, Khosla, Bernstein, Berg, and Fei-Fei}{Russakovsky
  et~al\mbox{.}}{2015}]%
        {Russakovsky2015}
\bibfield{author}{\bibinfo{person}{Olga Russakovsky}, \bibinfo{person}{Jia
  Deng}, \bibinfo{person}{Hao Su}, \bibinfo{person}{Jonathan Krause},
  \bibinfo{person}{Sanjeev Satheesh}, \bibinfo{person}{Sean Ma},
  \bibinfo{person}{Zhiheng Huang}, \bibinfo{person}{Andrej Karpathy},
  \bibinfo{person}{Aditya Khosla}, \bibinfo{person}{Michael Bernstein},
  \bibinfo{person}{Alexander~C. Berg}, {and} \bibinfo{person}{Li Fei-Fei}.}
  \bibinfo{year}{2015}\natexlab{}.
\newblock \showarticletitle{ImageNet Large Scale Visual Recognition Challenge}.
\newblock \bibinfo{journal}{\emph{International Journal of Computer Vision}}
  \bibinfo{volume}{115}, \bibinfo{number}{3} (\bibinfo{date}{01 Dec}
  \bibinfo{year}{2015}), \bibinfo{pages}{211--252}.
\newblock
\showISSN{1573-1405}
\urldef\tempurl%
\url{https://doi.org/10.1007/s11263-015-0816-y}
\showDOI{\tempurl}


\bibitem[\protect\citeauthoryear{Said, Fields, Jain, and Albayrak}{Said
  et~al\mbox{.}}{2013}]%
        {DBLP:conf/cscw/SaidFJA13}
\bibfield{author}{\bibinfo{person}{Alan Said}, \bibinfo{person}{Ben Fields},
  \bibinfo{person}{Brijnesh~J. Jain}, {and} \bibinfo{person}{Sahin Albayrak}.}
  \bibinfo{year}{2013}\natexlab{}.
\newblock \showarticletitle{User-centric evaluation of a K-furthest neighbor
  collaborative filtering recommender algorithm}. In
  \bibinfo{booktitle}{\emph{Proc. {CSCW}}}. \bibinfo{publisher}{{ACM}},
  \bibinfo{address}{San Antonio, TX, USA}, \bibinfo{pages}{1399--1408}.
\newblock


\bibitem[\protect\citeauthoryear{Schoeffmann, Bailer, Gurrin, Awad, and
  Loko\v{c}}{Schoeffmann et~al\mbox{.}}{2018}]%
        {Schoeffmann:2018:IVS:3240508.3241473}
\bibfield{author}{\bibinfo{person}{Klaus Schoeffmann}, \bibinfo{person}{Werner
  Bailer}, \bibinfo{person}{Cathal Gurrin}, \bibinfo{person}{George Awad},
  {and} \bibinfo{person}{Jakub Loko\v{c}}.} \bibinfo{year}{2018}\natexlab{}.
\newblock \showarticletitle{Interactive Video Search: Where is the User in the
  Age of Deep Learning?}. In \bibinfo{booktitle}{\emph{Proc ACM Multimedia}}.
  \bibinfo{publisher}{ACM}, \bibinfo{address}{Seoul, Republic of Korea},
  \bibinfo{pages}{2101--2103}.
\newblock


\bibitem[\protect\citeauthoryear{Shakhnarovich, Darrell, and
  Indyk}{Shakhnarovich et~al\mbox{.}}{2006}]%
        {SDI06}
\bibfield{author}{\bibinfo{person}{Gregory Shakhnarovich},
  \bibinfo{person}{Trevor Darrell}, {and} \bibinfo{person}{Piotr Indyk}.}
  \bibinfo{year}{2006}\natexlab{}.
\newblock \showarticletitle{Nearest-Neighbor Methods in Learning and Vision:
  Theory and Practice}.
\newblock \bibinfo{journal}{\emph{Pattern Analysis and Applications}}
  \bibinfo{volume}{11}, \bibinfo{number}{2} (\bibinfo{year}{2006}).
\newblock


\bibitem[\protect\citeauthoryear{Sigur{\dh}ard\'ottir, Hauksson, J\'{o}nsson,
  and Amsaleg}{Sigur{\dh}ard\'ottir et~al\mbox{.}}{2005}]%
        {EMMA}
\bibfield{author}{\bibinfo{person}{Rut Sigur{\dh}ard\'ottir},
  \bibinfo{person}{Hlynur Hauksson}, \bibinfo{person}{Bj\"{o}rn~{\TH}\'or
  J\'{o}nsson}, {and} \bibinfo{person}{Laurent Amsaleg}.}
  \bibinfo{year}{2005}\natexlab{}.
\newblock \showarticletitle{The Quality vs. Time Tradeoff for Approximate Image
  Descriptor Search}. In \bibinfo{booktitle}{\emph{Proc. IEEE EMMA workshop}}.
  \bibinfo{publisher}{IEEE}, \bibinfo{address}{Tokyo, Japan}.
\newblock


\bibitem[\protect\citeauthoryear{Snoek, Worring, de~Rooij, van~de Sande, Yan,
  and Hauptmann}{Snoek et~al\mbox{.}}{2008}]%
        {videolympics}
\bibfield{author}{\bibinfo{person}{C.G.M. Snoek}, \bibinfo{person}{M. Worring},
  \bibinfo{person}{O. de Rooij}, \bibinfo{person}{K.E.A. van~de Sande},
  \bibinfo{person}{Rong Yan}, {and} \bibinfo{person}{A.G. Hauptmann}.}
  \bibinfo{year}{2008}\natexlab{}.
\newblock \showarticletitle{VideOlympics: Real-Time Evaluation of Multimedia
  Retrieval Systems}.
\newblock \bibinfo{journal}{\emph{IEEE MM}} \bibinfo{volume}{15},
  \bibinfo{number}{1} (\bibinfo{year}{2008}), \bibinfo{pages}{86--91}.
\newblock
\showISSN{1070-986X}
\urldef\tempurl%
\url{https://doi.org/10.1109/MMUL.2008.21}
\showDOI{\tempurl}


\bibitem[\protect\citeauthoryear{Szegedy, Liu, Jia, Sermanet, Reed, Anguelov,
  Erhan, Vanhoucke, and Rabinovich}{Szegedy et~al\mbox{.}}{2015}]%
        {7298594}
\bibfield{author}{\bibinfo{person}{C. Szegedy}, \bibinfo{person}{Wei Liu},
  \bibinfo{person}{Yangqing Jia}, \bibinfo{person}{P. Sermanet},
  \bibinfo{person}{S. Reed}, \bibinfo{person}{D. Anguelov}, \bibinfo{person}{D.
  Erhan}, \bibinfo{person}{V. Vanhoucke}, {and} \bibinfo{person}{A.
  Rabinovich}.} \bibinfo{year}{2015}\natexlab{}.
\newblock \showarticletitle{Going deeper with convolutions}. In
  \bibinfo{booktitle}{\emph{Proc. IEEE CVPR}}. \bibinfo{publisher}{IEEE
  Computer Society}, \bibinfo{address}{Boston, MA, USA}, \bibinfo{pages}{1--9}.
\newblock


\bibitem[\protect\citeauthoryear{Tao, Yi, Sheng, and Kalnis}{Tao
  et~al\mbox{.}}{2009}]%
        {Tao:2009:QEH:1559845.1559905}
\bibfield{author}{\bibinfo{person}{Yufei Tao}, \bibinfo{person}{Ke Yi},
  \bibinfo{person}{Cheng Sheng}, {and} \bibinfo{person}{Panos Kalnis}.}
  \bibinfo{year}{2009}\natexlab{}.
\newblock \showarticletitle{Quality and Efficiency in High Dimensional Nearest
  Neighbor Search}. In \bibinfo{booktitle}{\emph{Proceedings of the ACM
  International Conference on Management of Data}}. \bibinfo{publisher}{ACM},
  \bibinfo{address}{Boston, MA, USA}, \bibinfo{pages}{563--576}.
\newblock


\bibitem[\protect\citeauthoryear{Tavenard, J{\'e}gou, and Amsaleg}{Tavenard
  et~al\mbox{.}}{2011}]%
        {tavenard:inria-00576886}
\bibfield{author}{\bibinfo{person}{Romain Tavenard}, \bibinfo{person}{Herv{\'e}
  J{\'e}gou}, {and} \bibinfo{person}{Laurent Amsaleg}.}
  \bibinfo{year}{2011}\natexlab{}.
\newblock \showarticletitle{{Balancing clusters to reduce response time
  variability in large scale image search}}. In
  \bibinfo{booktitle}{\emph{International Workshop on Content-Based Multimedia
  Indexing}}. \bibinfo{publisher}{IEEE}, \bibinfo{address}{Madrid, Spain}.
\newblock


\bibitem[\protect\citeauthoryear{Thomee, Shamma, Friedland, Elizalde, Ni,
  Poland, Borth, and Li}{Thomee et~al\mbox{.}}{2016}]%
        {Thomee:2016:YND:2886013.2812802}
\bibfield{author}{\bibinfo{person}{Bart Thomee}, \bibinfo{person}{David~A.
  Shamma}, \bibinfo{person}{Gerald Friedland}, \bibinfo{person}{Benjamin
  Elizalde}, \bibinfo{person}{Karl Ni}, \bibinfo{person}{Douglas Poland},
  \bibinfo{person}{Damian Borth}, {and} \bibinfo{person}{Li-Jia Li}.}
  \bibinfo{year}{2016}\natexlab{}.
\newblock \showarticletitle{{YFCC100M}: The New Data in Multimedia Research}.
\newblock \bibinfo{journal}{\emph{Commun. ACM}} \bibinfo{volume}{59},
  \bibinfo{number}{2} (\bibinfo{year}{2016}), \bibinfo{pages}{64--73}.
\newblock


\bibitem[\protect\citeauthoryear{Vijayanarasimhan, Jain, and
  Grauman}{Vijayanarasimhan et~al\mbox{.}}{2014}]%
        {DBLP:journals/pami/Vijayanarasimhan0G14}
\bibfield{author}{\bibinfo{person}{Sudheendra Vijayanarasimhan},
  \bibinfo{person}{Prateek Jain}, {and} \bibinfo{person}{Kristen Grauman}.}
  \bibinfo{year}{2014}\natexlab{}.
\newblock \showarticletitle{Hashing Hyperplane Queries to Near Points with
  Applications to Large-Scale Active Learning}.
\newblock \bibinfo{journal}{\emph{{IEEE} Trans. Pattern Anal. Mach. Intell.}}
  \bibinfo{volume}{36}, \bibinfo{number}{2} (\bibinfo{year}{2014}),
  \bibinfo{pages}{276--288}.
\newblock


\bibitem[\protect\citeauthoryear{Wang, Kumar, and Chang}{Wang
  et~al\mbox{.}}{2010}]%
        {DBLP:conf/cvpr/WangKC10}
\bibfield{author}{\bibinfo{person}{Jun Wang}, \bibinfo{person}{Ondrej Kumar},
  {and} \bibinfo{person}{Shih{-}Fu Chang}.} \bibinfo{year}{2010}\natexlab{}.
\newblock \showarticletitle{Semi-supervised hashing for scalable image
  retrieval}. In \bibinfo{booktitle}{\emph{Proceedings of the IEEE
  International Conference on Computer Vision \& Pattern Recognition}}.
  \bibinfo{publisher}{IEEE Computer Society}, \bibinfo{address}{San Francisco,
  CA, USA}, \bibinfo{pages}{3424--3431}.
\newblock


\bibitem[\protect\citeauthoryear{{Wang}, {Zhang}, j.~{song}, {Sebe}, and
  {Shen}}{{Wang} et~al\mbox{.}}{2018}]%
        {7915742}
\bibfield{author}{\bibinfo{person}{J. {Wang}}, \bibinfo{person}{T. {Zhang}},
  \bibinfo{person}{j. {song}}, \bibinfo{person}{N. {Sebe}}, {and}
  \bibinfo{person}{H.~T. {Shen}}.} \bibinfo{year}{2018}\natexlab{}.
\newblock \showarticletitle{A Survey on Learning to Hash}.
\newblock \bibinfo{journal}{\emph{IEEE Transactions on Pattern Analysis and
  Machine Intelligence}} \bibinfo{volume}{40}, \bibinfo{number}{4}
  (\bibinfo{date}{April} \bibinfo{year}{2018}), \bibinfo{pages}{769--790}.
\newblock
\showISSN{0162-8828}
\urldef\tempurl%
\url{https://doi.org/10.1109/TPAMI.2017.2699960}
\showDOI{\tempurl}


\bibitem[\protect\citeauthoryear{Wang, Zhang, Li, Zhang, and Lin}{Wang
  et~al\mbox{.}}{2017}]%
        {Wang:2017:CAL:3203306.3203314}
\bibfield{author}{\bibinfo{person}{Keze Wang}, \bibinfo{person}{Dongyu Zhang},
  \bibinfo{person}{Ya Li}, \bibinfo{person}{Ruimao Zhang}, {and}
  \bibinfo{person}{Liang Lin}.} \bibinfo{year}{2017}\natexlab{}.
\newblock \showarticletitle{Cost-Effective Active Learning for Deep Image
  Classification}.
\newblock \bibinfo{journal}{\emph{IEEE Trans. Cir. and Sys. for Video
  Technol.}} \bibinfo{volume}{27}, \bibinfo{number}{12} (\bibinfo{year}{2017}),
  \bibinfo{pages}{2591--2600}.
\newblock


\bibitem[\protect\citeauthoryear{Weiss, Torralba, and Fergus}{Weiss
  et~al\mbox{.}}{2008}]%
        {DBLP:conf/nips/WeissTF08}
\bibfield{author}{\bibinfo{person}{Yair Weiss}, \bibinfo{person}{Antonio
  Torralba}, {and} \bibinfo{person}{Robert Fergus}.}
  \bibinfo{year}{2008}\natexlab{}.
\newblock \showarticletitle{Spectral Hashing}. In
  \bibinfo{booktitle}{\emph{Neural Information Processing Systems}}.
  \bibinfo{publisher}{Curran Associates}, \bibinfo{address}{Vancouver, BC,
  Canada}, \bibinfo{pages}{1753--1760}.
\newblock


\bibitem[\protect\citeauthoryear{Williams}{Williams}{2004}]%
        {DBLP:conf/icalp/Williams04}
\bibfield{author}{\bibinfo{person}{Ryan Williams}.}
  \bibinfo{year}{2004}\natexlab{}.
\newblock \showarticletitle{A New Algorithm for Optimal Constraint Satisfaction
  and Its Implications}. In \bibinfo{booktitle}{\emph{Proc. {ICALP}}}.
  \bibinfo{publisher}{Springer}, \bibinfo{address}{Turku, Finland},
  \bibinfo{pages}{1227--1237}.
\newblock


\bibitem[\protect\citeauthoryear{Xioufis, Papadopoulos, Kompatsiaris,
  Tsoumakas, and Vlahavas}{Xioufis et~al\mbox{.}}{2014}]%
        {DBLP:journals/tmm/XioufisPKTV14}
\bibfield{author}{\bibinfo{person}{Eleftherios~Spyromitros Xioufis},
  \bibinfo{person}{Symeon Papadopoulos}, \bibinfo{person}{Yiannis
  Kompatsiaris}, \bibinfo{person}{Grigorios Tsoumakas}, {and}
  \bibinfo{person}{Ioannis~P. Vlahavas}.} \bibinfo{year}{2014}\natexlab{}.
\newblock \showarticletitle{A Comprehensive Study Over {VLAD} and Product
  Quantization in Large-Scale Image Retrieval}.
\newblock \bibinfo{journal}{\emph{IEEE Transactions on Multimedia}}
  \bibinfo{volume}{16}, \bibinfo{number}{6} (\bibinfo{year}{2014}),
  \bibinfo{pages}{1713--1728}.
\newblock


\bibitem[\protect\citeauthoryear{Xu, Bao, Yao, Zhou, Tang, Guo, and Xu}{Xu
  et~al\mbox{.}}{2017}]%
        {DBLP:journals/jcst/XuBYZTGX17}
\bibfield{author}{\bibinfo{person}{Xiao{-}Jun Xu}, \bibinfo{person}{Jin{-}Song
  Bao}, \bibinfo{person}{Bin Yao}, \bibinfo{person}{Jingyu Zhou},
  \bibinfo{person}{Feilong Tang}, \bibinfo{person}{Minyi Guo}, {and}
  \bibinfo{person}{Jianqiu Xu}.} \bibinfo{year}{2017}\natexlab{}.
\newblock \showarticletitle{Reverse Furthest Neighbors Query in Road Networks}.
\newblock \bibinfo{journal}{\emph{J. Comput. Sci. Technol.}}
  \bibinfo{volume}{32}, \bibinfo{number}{1} (\bibinfo{year}{2017}),
  \bibinfo{pages}{155--167}.
\newblock


\bibitem[\protect\citeauthoryear{Yang, Ma, Nie, Chang, and Hauptmann}{Yang
  et~al\mbox{.}}{2015}]%
        {Yang2015}
\bibfield{author}{\bibinfo{person}{Yi Yang}, \bibinfo{person}{Zhigang Ma},
  \bibinfo{person}{Feiping Nie}, \bibinfo{person}{Xiaojun Chang}, {and}
  \bibinfo{person}{Alexander~G. Hauptmann}.} \bibinfo{year}{2015}\natexlab{}.
\newblock \showarticletitle{Multi-Class Active Learning by Uncertainty Sampling
  with Diversity Maximization}.
\newblock \bibinfo{journal}{\emph{International Journal of Computer Vision}}
  \bibinfo{volume}{113}, \bibinfo{number}{2} (\bibinfo{date}{01 Jun}
  \bibinfo{year}{2015}), \bibinfo{pages}{113--127}.
\newblock
\showISSN{1573-1405}
\urldef\tempurl%
\url{https://doi.org/10.1007/s11263-014-0781-x}
\showDOI{\tempurl}


\bibitem[\protect\citeauthoryear{Zah\'{a}lka, Rudinac, and Worring}{Zah\'{a}lka
  et~al\mbox{.}}{2015}]%
        {Zahalka:2015:AQE:2733373.2806279}
\bibfield{author}{\bibinfo{person}{Jan Zah\'{a}lka}, \bibinfo{person}{Stevan
  Rudinac}, {and} \bibinfo{person}{Marcel Worring}.}
  \bibinfo{year}{2015}\natexlab{}.
\newblock \showarticletitle{Analytic Quality: Evaluation of Performance and
  Insight in Multimedia Collection Analysis}. In
  \bibinfo{booktitle}{\emph{Proc. ACM Multimedia}}. \bibinfo{publisher}{ACM},
  \bibinfo{address}{Brisbane, Australia}, \bibinfo{pages}{231--240}.
\newblock


\bibitem[\protect\citeauthoryear{Zahálka, Rudinac, Jónsson, Koelma, and
  Worring}{Zahálka et~al\mbox{.}}{2018a}]%
        {blackthorn_tmm}
\bibfield{author}{\bibinfo{person}{J. Zahálka}, \bibinfo{person}{S. Rudinac},
  \bibinfo{person}{B.~Þ. Jónsson}, \bibinfo{person}{D.~C. Koelma}, {and}
  \bibinfo{person}{M. Worring}.} \bibinfo{year}{2018}\natexlab{a}.
\newblock \showarticletitle{Blackthorn: Large-Scale Interactive Multimodal
  Learning}.
\newblock \bibinfo{journal}{\emph{IEEE Transactions on Multimedia}}
  \bibinfo{volume}{20}, \bibinfo{number}{3} (\bibinfo{date}{March}
  \bibinfo{year}{2018}), \bibinfo{pages}{687--698}.
\newblock
\showISSN{1520-9210}
\urldef\tempurl%
\url{https://doi.org/10.1109/TMM.2017.2755986}
\showDOI{\tempurl}


\bibitem[\protect\citeauthoryear{Zahálka, Rudinac, Jónsson, Koelma, and
  Worring}{Zahálka et~al\mbox{.}}{2018b}]%
        {8048559}
\bibfield{author}{\bibinfo{person}{J. Zahálka}, \bibinfo{person}{S. Rudinac},
  \bibinfo{person}{B.~Þ. Jónsson}, \bibinfo{person}{D.~C. Koelma}, {and}
  \bibinfo{person}{M. Worring}.} \bibinfo{year}{2018}\natexlab{b}.
\newblock \showarticletitle{Blackthorn: Large-Scale Interactive Multimodal
  Learning}.
\newblock \bibinfo{journal}{\emph{IEEE Transactions on Multimedia}}
  \bibinfo{volume}{20}, \bibinfo{number}{3} (\bibinfo{date}{March}
  \bibinfo{year}{2018}), \bibinfo{pages}{687--698}.
\newblock
\showISSN{1520-9210}
\urldef\tempurl%
\url{https://doi.org/10.1109/TMM.2017.2755986}
\showDOI{\tempurl}


\bibitem[\protect\citeauthoryear{Zahálka and Worring}{Zahálka and
  Worring}{2014}]%
        {7042476}
\bibfield{author}{\bibinfo{person}{J. Zahálka} {and} \bibinfo{person}{M.
  Worring}.} \bibinfo{year}{2014}\natexlab{}.
\newblock \showarticletitle{Towards interactive, intelligent, and integrated
  multimedia analytics}. In \bibinfo{booktitle}{\emph{Proc. IEEE Conference on
  Visual Analytics Science and Technology (VAST)}}. \bibinfo{publisher}{IEEE
  Computer Society}, \bibinfo{address}{Paris, France}, \bibinfo{pages}{3--12}.
\newblock


\bibitem[\protect\citeauthoryear{Zhang, Agrawal, Chen, and Tung}{Zhang
  et~al\mbox{.}}{2011}]%
        {5767837}
\bibfield{author}{\bibinfo{person}{Dongxiang Zhang}, \bibinfo{person}{D.
  Agrawal}, \bibinfo{person}{Gang Chen}, {and} \bibinfo{person}{A. Tung}.}
  \bibinfo{year}{2011}\natexlab{}.
\newblock \showarticletitle{HashFile: An efficient index structure for
  multimedia data}. In \bibinfo{booktitle}{\emph{Proceedings of the IEEE
  International Conference on Data Engineering}}. \bibinfo{publisher}{IEEE
  Computer Society}, \bibinfo{address}{Hannover, Germany},
  \bibinfo{pages}{1103--1114}.
\newblock


\end{thebibliography}

\end{document}